\documentstyle[emulateapj,onecolfloat]{article}
\newcommand{\mpc}{{\rm\,Mpc}}
\newcommand{\impc}{{\rm\,Mpc}^{-1}}
\newcommand{\hmpc}{h^{-1}{\rm\,Mpc}}
\newcommand{\ihmpc}{h{\rm\,Mpc}^{-1}}
\newcommand{\kmsmpc}{{\rm\ km\ s^{-1}\ Mpc^{-1}}}
\newcommand{\eV}{{\rm\ eV}}
\newcommand{\om}{\Omega_m}
\newcommand{\omhh}{\om h^2}
\newcommand{\ob}{\Omega_B}
\newcommand{\obhh}{\ob h^2}
\newcommand{\ol}{\Omega_\Lambda}
\newcommand{\ok}{\Omega_K}
\newcommand{\on}{\Omega_\nu}
\newcommand{\da}{d_A}
\newcommand{\ns}{n_S}
\newcommand{\nt}{n_T}
\newcommand{\ts}{T/S}
\newcommand{\As}{A^2_S}
\newcommand{\bfw}{{\mathbf{w}}}
\newcommand{\bfx}{{\mathbf{x}}}
\newcommand{\bfr}{{\mathbf{r}}}
\newcommand{\bfF}{{\mathbf{F}}}
\newcommand{\bfp}{{\mathbf{p}}}
\newcommand{\Cov}{{\rm Cov}}
\newcommand{\kmax}{k_{\rm max}}
\newcommand{\kfid}{k_{\rm fid}}
\newcommand{\kpivot}{k_{\rm pivot}}
\newcommand{\knorm}{k_{\rm norm}}
\newcommand{\Pinit}{P_{\Phi}}
\newcommand{\fsky}{f_{\rm sky}}
\newcommand{\dgr}{D_{\rm gr}}
\newcommand{\tableskip}{\tablevspace{3pt}}
\newlength{\tskip}
\newlength{\colwidth}
\newcommand{\colskip}{@{\hspace{0.3in}}}
\newcommand{\etal}{et al.}

\newcommand{\beq}{\begin{equation}}
\newcommand{\eeq}{\end{equation}}
\newcommand{\beqa}{\begin{eqnarray}}
\newcommand{\eeqa}{\end{eqnarray}}
\newcommand{\beam}{B_\ell}	
\newcommand{\nbar}{\bar{n}}	
\newcommand{\Dp}{\Delta p}
\newcommand{\fpt}{\tilde{f}'}
\def\simlt{\lesssim}

\begin{document}
\twocolumn[
\submitted{Submitted to ApJ July 2, 1998}
\title{Cosmic Complementarity: Joint Parameter Estimation from CMB Experiments and Redshift Surveys}
\author{Daniel J. Eisenstein, Wayne Hu\altaffilmark{1}, and Max Tegmark\altaffilmark{2}}
\affil{Institute for Advanced Study, Princeton, NJ 08540}

\begin{abstract}
We study the ability of future CMB anisotropy experiments and 
redshift surveys to constrain a thirteen-dimensional parameterization
of the adiabatic cold dark matter model.  
Each alone is unable to 
determine all parameters to high accuracy.  However, considered
together, one data set resolves the difficulties of the other,
allowing certain degenerate parameters to be determined with far
greater precision.  We treat in detail the degeneracies involving the classical
cosmological parameters, massive neutrinos, tensor-scalar ratio, 
bias, and reionization optical depth as well as
how redshift surveys can resolve them. 
We discuss the opportunities for internal and external 
consistency checks on these measurements.
Previous papers on parameter estimation have generally treated 
smaller parameter spaces;
in direct comparisons to these works, we tend to find weaker constraints
and suggest numerical explanations for the discrepancies.
\end{abstract}

\keywords{cosmology: theory -- dark matter -- large-scale structure of 
the universe -- cosmic microwave background}
]
\altaffiltext{1}{Alfred P. Sloan Fellow}
\altaffiltext{2}{Hubble Fellow}

\section{Introduction}\label{sec:intro}
Current cosmological data have led astrophysicists to explore a
structure formation paradigm in which cold dark matter driven by
adiabatic fluctuations leads to the formation of galaxies and CMB
anisotropies (see 
\cite{Blu84} 1984;
\cite{Dod96b} 1996 for reviews).  
Experiments planned for the next decade will be
able to test this paradigm more stringently by searching for 
distinctive features in the power spectra of CMB anisotropies
and polarization (see \cite{Hu97}\ 1997 and \cite{Hu97p}\ 1997b
for reviews).  
If this framework is confirmed, then upcoming measurements, notably
from the MAP\footnote{http://map.gsfc.nasa.gov} and 
Planck\footnote{http://astro.estec.esa.nl/SA-general/Projects/Planck}
satellites, will enable
precision measurements of cosmological parameters such as the baryon
fraction and matter-radiation ratio
(\cite{Jun96a}\ 1996a,b; \cite{Zal97a}\ 1997; \cite{Bon97}\ 1997;
\cite{Cop98}\ 1998; \cite{Sto98}\ 1998).

As discussed by a number of authors, the details of the CMB 
power spectra contain a considerable amount of cosmological 
information.  However, this leverage is not complete; altering
the model parameters in particular combinations can yield power
spectra that are observationally indistinguishable from a reference model
(\cite{Bon94}\ 1994, 1997; \cite{Zal97a} 1997).
The presence of these so-called degenerate directions means that a CMB data
set will restrict the allowed models to a curve or surface in 
parameter space rather a point.  

The galaxy power spectrum of large redshift surveys such as the 
2dF survey\footnote{http://meteor.anu.edu.au/$\sim$colless/2dF} and the 
Sloan Digital Sky Survey (SDSS)\footnote{http://www.astro.princeton.edu/BBOOK}
provide a different 
window on the cosmological parameter space.  Taken alone, the
results are again plagued by degeneracies 
(\cite{Teg97a} 1997a; \cite{Gol98} 1998; Hu \etal\ 1998, hereafter \cite{HET}).
However,
when combined with CMB data, the constraints can be highly
complementary in that the degenerate directions of one lie along
well-constrained directions for the other.
A striking example involves the Hubble constant. 
By combining the data sets, features in the power spectrum
can be measured in both real and redshift space, allowing the
Hubble constant to be identified even though neither data set
alone provides a good constraint
(Eisenstein \etal\ 1998, hereafter \cite{EHT}).  

In this paper, we explore the details of how these data sets
complement each other.  In particular, we consider the
constraints on cosmological parameters
attainable by the statistical errors of this next generation of
CMB experiments and redshift surveys.  We identify physical
mechanisms by which the data sets resolve degeneracies
and explore them through progressions of cosmological models 
in the baryon fraction, neutrino mass, and tensor contribution.
The combination of CMB data and large-scale structure has been studied
with current data (\cite{Sco95}\ 1995; \cite{Bon96}\ 1996;
\cite{Lin98}\ 1998; \cite{Gaw98} 1998; \cite{Web98}\ 1998) 
as well as with future data in a
smaller space of cosmological parameters but with more
general initial conditions (\cite{Wan98}\ 1998).
We study how reducing the cosmological parameter space
affects the degeneracies.

In addition to incorporating large-scale structure data, our treatment
of parameter estimation in the case of CMB data alone uses the most
general cosmology yet studied with both temperature and polarization
information.  Taking account of differences in cosmological
parameterizations, fiducial models, and experimental specifications, we
compare our results with past work in a series of tables.  We generally
find stronger degeneracies and hence weaker constraints than previous
papers and propose numerical explanations for the discrepancies.

We review parameter estimation methods in \S\,\ref{sec:fisher}
and describe our parameterization of cosmology in
\S\,\ref{sec:parameters}.  
We present a way of interpreting
parameter covariance in Appendix \ref{sec:minerror}
and discuss the need for careful numerical
treatments in Appendix \ref{sec:numerics}.  
In \S\,\ref{sec:cmbalone}, we compute the precision with which upcoming CMB
experiments can potentially measure cosmological parameters.  
We compare our
results to previous studies in Appendix \ref{sec:comparison}.  
In \S\,\ref{sec:comp}, we add redshift survey information and
conduct several parameter studies.
We explore the dependence of our results on our assumptions in 
\S \ref{sec:assumptions}.
In \S\,\ref{sec:consist}, we discuss the necessity of cosmological
consistency checks and highlight a number of possibilities.  We
conclude in \S\,\ref{sec:concl}.

\section{Fisher Matrix Methods}\label{sec:fisher}

The Fisher information matrix encodes the manner in which experimental
data depends upon a set of underlying theoretical parameters that
one wishes to measure (see \cite{Teg97b}\ 1997b for a review).  
Within this set of parameters, the Fisher
matrix yields a lower limit to error bars and hence an upper limit
on the information that can be extracted from such a data set.
With the further assumptions of Gaussian-distributed signal
and noise, Fisher matrices can be 
constructed from the specifications of both CMB experiments 
(\cite{Jun96b}\ 1996b; \cite{Sel96b}\ 1996b; \cite{Zal97b}\ 1997; 
\cite{Kam97}\ 1997)
and redshift surveys (\cite{Teg97a}\ 1997a).  
To the extent that the data sets are independent---a very good 
approximation for the cosmologies of interest---we can combine their 
constraints simply by summing their Fisher matrices.  

Suppose that the observed data are written as $x_1$, $x_2$, $\ldots x_n$,
arranged as a vector $\bfx$.  Then suppose that the 
model parameters are $p_1$, $p_2$, $\ldots p_m$, 
arranged as a vector $\bfp$.
Let the probability of observing a set of data $\bfx$ given the
true parameters $\bfp$ (the ``fiducial model'') be $L(\bfx; \bfp)$.
The Fisher matrix is then defined as 
\beq\label{eq:fisherdef}
\bfF_{ij} = - \left<\partial^2 \ln L\over 
	\partial p_i \partial p_j\right>_\bfx.
\eeq
The Cram\'er-Rao inequality says that the variance 
of an unbiased estimator
of a parameter $p_i$ from a data set cannot be less than
$(\bfF^{-1})_{ii}$.  In this sense, the Fisher matrix reveals
the best possible statistical error bars achievable from an
experiment.

Since the error bars on our independent variables $\bfp$ will
in general be correlated, they do not contain sufficient information
to calculate the errors on constructed quantities.   
A change of variables shows that the best possible error bar on
some quantity $g(\bfp)$ when marginalizing over the other 
directions that span the parameter space is
\beq\label{eq:transform}
\sigma^2_g = \sum_{i,j} \left(\partial g\over\partial p_i\right)
	(\bfF^{-1})_{ij} \left(\partial g\over\partial p_j\right).
\eeq
More generally, the Fisher matrix transforms as a tensor
under a change of variables in parameter space.
We will use equation (\ref{eq:transform}) to quote errors for a number of these
constructed quantities, such as $\sigma_R$, the rms mass fluctuations
in a sphere of $R\hmpc$ radius.

It is at times convenient to think of the inverse of the Fisher matrix
as a covariance matrix with an associated error ellipsoid.  This view
can be misleading.  First, it represents a degeneracy as a straight
line rather than the true curve.  For example, a CMB experiment might
determine $\omhh$ well but neither $\om$ nor $h$ well.  The proper
error contour in the $\om$--$h$ plane would be a banana-shaped region
along a curve of constant $\omhh$; however, it will instead be
represented as a long ellipse with the slope of the $\omhh$ curve at
the location of the fiducial model.  

Second, the error contours from
the Fisher matrix are not necessarily those that would be obtained from
a likelihood or goodness-of-fit analysis of a particular data set.  
If the Fisher matrix errors are roughly
constant across the error region itself (i.e. if the likelihood
function is nearly Gaussian), then these various error bars will be
comparable.  This generally occurs when the error
estimates are small compared to characteristic range over which a given
cosmological parameter affects model predictions
(\cite{Zal97a}\ 1997a).  As we will see, this is usually the case when
CMB and redshift surveys are combined, but CMB data alone is subject to
degenerate directions that surely violate the approximation.  
In these cases, the question of what confidence region to use descends
into the murky debate between frequentists and Bayesians.
If a strong degeneracy is present, the Fisher matrix method
is guaranteed to find it, but different methods may disagree on
the size and shape of the error region.
After this paper was submitted, a paper investigating this issue
with Monte Carlo methods was submitted by \cite{Efs98} (1998).

\subsection{CMB anisotropies}\label{sec:cmbfisher}

Under the assumption of Gaussian perturbations and Gaussian noise,
the Fisher matrix for CMB anisotropies and polarization is
(\cite{Sel96b}\ 1996b; \cite{Zal97b}\ 1997; \cite{Kam97}\ 1997)
\beq\label{eq:fishercmb}
\bfF_{ij} = \sum_\ell\sum_{X,Y} {\partial C_{X\ell}\over\partial p_i}
(\Cov_\ell)^{-1}_{XY}{\partial C_{X\ell}\over\partial p_j},
\eeq
where $C_{X\ell}$ is the power in the $\ell^{th}$ multipole for 
$X=T$, $E$, $B$, and $C$---the temperature, $E$-channel polarization,
$B$-channel polarization, and temperature-polarization cross-correlation,
respectively.  We will use $C_\ell$ at times to refer to all the CMB
power spectra together.  The elements of the (symmetric) covariance matrix 
$\Cov_\ell$ between the various power spectra are
\beqa\label{eq:cmbcov}
(\Cov_\ell)_{TT}&=&{2\over(2\ell+1)\fsky}(C_{T\ell}+w_T^{-1}\beam^{-2})^2, \\
(\Cov_\ell)_{EE}&=&{2\over(2\ell+1)\fsky}(C_{E\ell}+w_P^{-1}\beam^{-2})^2, \\
(\Cov_\ell)_{BB}&=&{2\over(2\ell+1)\fsky}(C_{B\ell}+w_P^{-1}\beam^{-2})^2, \\
(\Cov_\ell)_{CC}&=&{1\over(2\ell+1)\fsky}\left[C_{C\ell}^2
    +(C_{T\ell}+w_T^{-1}\beam^{-2})
    \right.\nonumber\\&&\left.
    \times(C_{E\ell}+w_P^{-1}\beam^{-2})\right],\\
(\Cov_\ell)_{TE}&=&{2\over(2\ell+1)\fsky}C_{C\ell}^2, \\
(\Cov_\ell)_{TC}&=&{2\over(2\ell+1)\fsky}C_{C\ell}
	(C_{T\ell}+w_T^{-1}\beam^{-2}), \\
(\Cov_\ell)_{EC}&=&{2\over(2\ell+1)\fsky}C_{C\ell}
	(C_{E\ell}+w_P^{-1}\beam^{-2}), \\
(\Cov_\ell)_{TB}&=&(\Cov_\ell)_{EB} = (\Cov_\ell)_{CB} = 0.
\eeqa
Here $\beam^2$ is the beam window function, assumed Gaussian with
$\beam^2=\exp(-\ell(\ell+1)\theta_{\rm beam}^2/8\ln2)$, where
$\theta_{\rm beam}$ is the full-width, half-maximum (FWHM) of the beam
in radians.  $w_T$ and $w_P$ are the inverse square of the detector
noise level on a steradian patch for temperature and polarization,
respectively.  A fully-polarized detector has $w_P=2w_T$.  
For multiple frequency channels, $w
\beam^2$ is replaced by the sum of this quantity for each channel.
These formulae are derived in the $\fsky=1$ case; 
the approximation for $\fsky<1$ including 
sample variance (\cite{Sco93}\ 1993) only gives the correct
Fisher matrix providing that the power spectra have no sharp 
spectral features on scales $\Delta\ell\simlt\Delta\theta^{-1}$,
where $\Delta\theta$ is the angular extent of the map
in the narrowest direction (\cite{Teg97b}\ 1997b).
Within the class of models we are considering, this 
should be an excellent approximation for MAP and Planck 
except at $\l\simlt 3$ where sample variance 
is large anyway.

We normalize the CMB power spectra to COBE when using
equation (\ref{eq:fishercmb}) (\cite{Bun97}\ 1997).

In Table \ref{tab:specs}, we list the experimental specifications
for the MAP and Planck satellites used in this paper.  
We use $\fsky=0.65$ in all cases.  
We are not using the 22 GHz or 30 GHz channels of MAP
and are using only 2 of the 10 channels of Planck.
The rationale is that the statistical power of these channels
will be used for multifrequency subtraction of foregrounds, leaving
the full power of the remaining channels for 
cosmological use.  We will discuss this further in \S\ \ref{sec:foregrounds}.

\begin{table}[tb]\footnotesize
\caption{\label{tab:specs}}
\begin{center}
{\sc CMB Experimental Specifications\\}
\begin{tabular}{rcccc}
\tableskip\tableline\tableline\tableskip
Experiment & Frequency & $\theta_{\rm beam}$ & 
$\sigma_T$ & $\sigma_P$ \\
\tableskip\tableline\tableskip
MAP: 
& 40 & 28.2 & 17.2 & 24.4 \\
& 60 & 21.0 & 30.0 & 42.6 \\
& 90 & 12.6 & 49.9 & 70.7 \\
\tableskip\tableline\tableskip
Planck: 
& 143 & 8.0 &  5.2 & 10.8 \\
& 217 & 5.5 &  11.7 & 24.3 \\
\tableskip\tableline
\end{tabular}
\end{center}
NOTES.---%
Frequencies in GHz.  Beam size $\theta_{\rm beam}$ is the FWHM in arcminutes.  
Sensitivities $\sigma_T$ and $\sigma_P$ are in $\mu$K 
per FWHM beam (and hence $C_\ell$ must be in $\mu$K$^2$).  
$w=(\theta_{\rm beam}\sigma)^{-2}$ is the weight
given to that channel.
\end{table}

\subsection{Redshift surveys}\label{sec:lssfisher}

For the power spectrum derived from galaxy redshift surveys, 
the Fisher matrix may be approximated as (\cite{Teg97a}\ 1997a)
\beqa\label{eq:fishlss}
\bfF_{ij}&=&\int_{k_{\rm min}}^{\kmax}
{\partial \ln P(k)\over \partial p_i}
{\partial \ln P(k)\over \partial p_j} V_{\rm eff}(k)
{k^2\,dk\over(2\pi)^2}, \\
 V_{\rm eff}(k)&=&\int \left[\nbar(\bfr)P(k)
\over 1+\nbar(\bfr)P(k)\right]^2 d^3r,
\label{eq:Veff}
\eeqa
where $\nbar(\bfr)$ is the survey selection function, i.e.\ the
expected number density of galaxies at the location $\bfr$.  $P(k)$ is
the model galaxy power spectrum.  $V_{\rm eff}(k)$ is the effective volume of
the survey, properly weighing the effects of shot noise in
undersampled regions.  $k_{\rm min}$ is the minimum wavenumber 
to which the survey is sensitive,
but in practice the numerical results are virtually 
unchanged by taking $k_{\rm min}=0$ since both
$V_{\rm eff}$ and the phase space factor $k^2$ vanish as
$k\to 0$.  $\kmax$ is the maximum wavenumber used for parameter
estimation; we use $0.1\ihmpc$ as a default but will discuss this at
length.

Equation (\ref{eq:fishlss}) 
was derived under the approximation that
the galaxy distribution is that of a Gaussian random field and that
the power spectrum has no features narrower than the inverse scale
of the survey.  It also neglects edge effects and redshift distortions.
Fortunately, the SDSS is both wide angle ($\pi$ steradians) and deep,
so power spectrum features (e.g.\ baryonic oscillations) should be
well-resolved and edge effects manageable 
(\cite{Hea97} 1997; \cite{THSVS} 1998b; c.f.\ \cite{Kai91} 1991).
The 2dF survey, however, has a more complicated geometry
and will require a more careful analysis.

We will quote our results for the Bright Red Galaxy (BRG) portion of
the SDSS.  This subsample will be intrinsically red galaxies; such
galaxies tend to be bright cluster galaxies and so the sample will
reach significantly deeper than the primary survey.  We assume the
sample to be volume-limited with $10^5$ galaxies to a depth of 1 Gpc
and to have a bias such that $\sigma_{8,{\rm gal}}=2$.
We will also consider the results for the main SDSS, which includes
$10^6$ galaxies with a more complicated radial selection function; we
assume this sample to have $\sigma_{8,{\rm gal}}=1$.  The value of
$\sigma_{8,{\rm gal}}$ affects the normalization of $P(k)$ in equation
(\ref{eq:Veff}); together with the assumed CMB normalization and
spectral tilt, it implies that the fiducial model may have
galaxy bias $b\neq 1$.

We will focus entirely on the power spectrum at large scales, where
linear theory is expected to be a good approximation.  We do this
by choosing $\kmax$ to be roughly the scale at which non-linear 
clustering becomes important (\cite{Teg97a}\ 1997a).  
While data on smaller scales will yield
very accurate measures of the power spectrum and higher-order
correlation functions, their interpretation in terms of cosmological
parameters is much more complicated
(\cite{Fry93} 1993; \cite{Pea97} 1997; \cite{Man98}\ 1998).  
To be conservative, we are neglecting the cosmological information
on non-linear scales.
SDSS should also reveal a wealth of information about redshift distortions
(e.g.\ \cite{Ham97} 1997; \cite{Hat98} 1998) on both linear and
non-linear scales; we will return to this in \S\,\ref{sec:consist}.

We assume that the galaxy bias is linear on these large scales. 
This has some theoretical justification 
(\cite{Col93}\ 1993; \cite{Fry93}\ 1993; \cite{Wei95}\ 1995; 
\cite{Sch98}\ 1997; \cite{Man98}\ 1998).  
More important,
this assumption will be stringently tested by the SDSS and other
large surveys.  Galaxies of different morphologies or type will
have different levels of bias, but linear bias predicts that 
the ratios of the various power spectra should be constant 
on large scales.  Scale-dependent bias has been detected on 
small scales (\cite{Pea97}\ 1997), but this does not test the linear
bias assumption we are making here.  In addition, redshift distortions 
may be able to probe any scale-dependence of bias on large scales.

\section{Parameterized Cosmology}\label{sec:parameters}

\begin{figure*}[t]
\centerline{\epsfxsize=\textwidth\epsffile{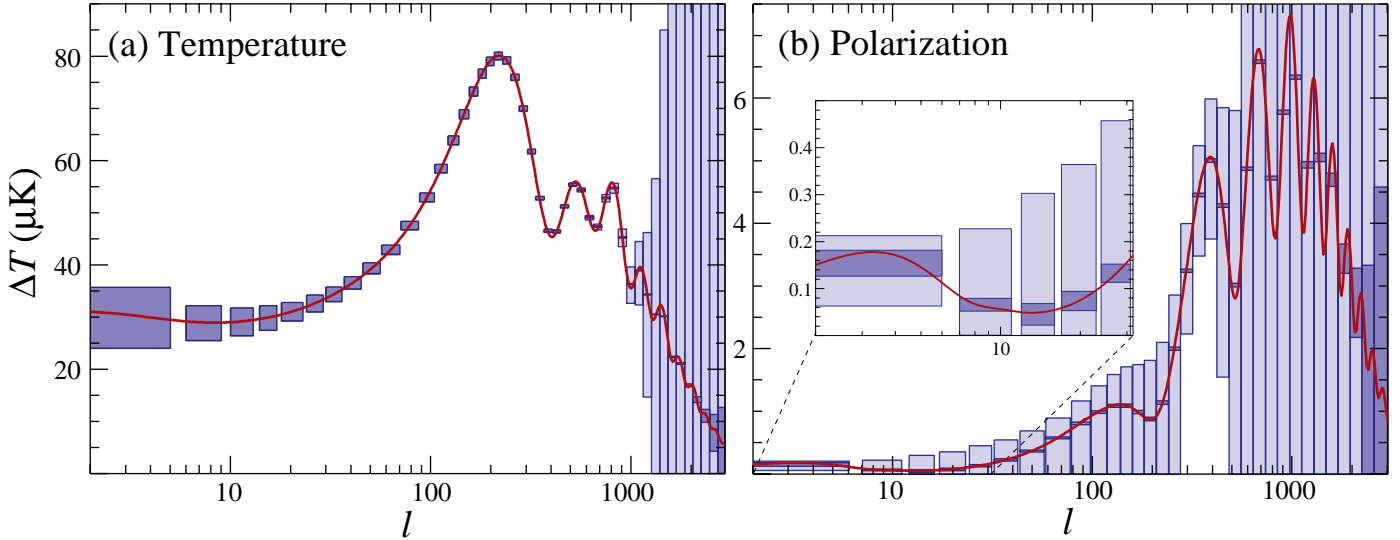}}
\caption{\label{fig:fidmodel}\footnotesize%
Our most used model along with 1--$\sigma$ band-power error bars from
MAP and Planck.  The model is $\om=0.35$, $h=0.65$, $\ob=0.05$,
$\ol=0.65$, $\on=0.0175$, $\tau=0.05$, $\ns(\kfid)=1$, and
$\ts=\alpha=0$.  a) The temperature power spectrum $\Delta T =
[\ell (\ell+1)C_{T\ell}/2\pi]^{1/2}$.  b) The
$E$-channel polarization power spectrum 
$\Delta T= [\ell (\ell+1)C_{E\ell}/2\pi]^{1/2}$; the panel shows a
blow-up of the large-angle feature caused by reionization.  MAP errors
are the lighter, larger boxes; Planck errors are the darker, smaller
boxes.  The bands reflect an averaging over many $\ell$; the actual
experiments will have finer $\ell$ resolution (and correspondingly
larger errors).  MAP will be able to average the polarization bands
together to get a marginal detection at $\ell\approx150$, but Planck
can trace out the full curve.  Note that because polarization and
temperature are correlated, the significance of detecting a change in
parameters using both data sets is not simply given by the combination of
errors from each.  
} \end{figure*}

We adopt a 13-dimensional parameterization of the adiabatic CDM model.
Our independent variables include the matter density $\om h^2$,
the baryon density $\ob h^2$, the massive neutrino density $\on h^2$,
the cosmological constant $\ol$, and a curvature contribution $\ok$.
Here, the Hubble constant is written as $H_0\equiv 100h\kmsmpc$.
These definitions imply that the total matter density in units of
the critical density $\om \equiv 1-\ol-\ok$,
that the Hubble constant $h\equiv\sqrt{(\om h^2)/\om}$, and that the
CDM density $\Omega_{\rm CDM}h^2\equiv \om h^2 - \ob h^2 - \on h^2$.
We use a single species of massive neutrinos, so the neutrino
mass is $m_\nu\approx94\on h^2 \eV$ 
(we take units in which the speed of light is unity).

We include an unknown optical depth $\tau$ to reionization, implemented
as a rapid and complete ionization event at the appropriate (small) redshift.
We also allow the primordial helium fraction to vary
but assume that the information from direct abundance measurements
can be represented by a Gaussian prior 
$Y_p=0.24\pm0.02$ (1-$\sigma$) (\cite{Bon97}\ 1997; \cite{Schr98} 1998).

We use an initial power spectrum of the form
\beq\label{eq:Pinit}
\Pinit = \As (k/\kfid)^{\ns(k)-4}
\eeq
with
\beq\label{eq:nk}
\ns(k) = \ns(\kfid) + \alpha \ln(k/\kfid)
\eeq
for the fluctuations in the gravitational potential.
Here, $\alpha$ is a logarithmic running
of the tilt around a fiducial scale $\kfid\equiv0.025\impc$.  
The density power spectrum is equal to the
potential power spectrum times $(k^2+3\ok H_0^2)^2$; 
the simplest open inflationary models predict a power-law in
the potential, not in the density.
In a flat universe, the initial density power spectrum takes on the usual 
$\As (k/\kfid)^{\ns}$ form.  The present-day density power spectrum 
$P(k)$ of course differs by the square of the transfer function.

Note that
equivalent parameters for a different choice of $\kfid$ could be mapped
into the parameters of equation (\ref{eq:Pinit}); hence the value of
$\kfid$ is immaterial.  However, because of the running of the tilt,
the value and hence the error bars on the tilt itself become scale-dependent.
We will quote values at both the Hubble wavenumber and at $\kfid$.
The errors on $\alpha$ are scale-independent
for the parameterization of equation (\ref{eq:Pinit}).
As shown in Appendix \ref{sec:minerror},
for a given experiment and fiducial model, there is a ``pivot''
wavenumber $\kpivot$ for
which the errors on $\ns(\kpivot)$ and $\alpha$ are uncorrelated and the errors
on $\ns(\kpivot)$ are equal to the errors on $\ns$ when $\alpha$ is held fixed.
As one might expect, $\kpivot$ falls near the center of the observable 
range of wavenumbers, generally not too far from our choice of $\kfid$.
At other wavenumbers, the uncertainties on $\ns(k)$ are larger and correlated
with $\alpha$.  

We allow tensor perturbations with a normalization $\ts$ equal to the
ratio of the $\ell=2$ temperature anisotropies of the tensors and scalars.
Note that this is proportional but not equal to the ratio $A_T^2/A_S^2$
that enters into the inflationary constraints. In particular,
cosmological parameters enter into $\ts$ due to the evolution of
quadrupole anisotropies (\cite{Kno95} 1995; \cite{Tur96} 1996).   
We parameterize the power-law exponent of the tensor input spectrum
as $\nt$; this allows us to probe whether these data sets can test
the inflationary consistency relation.  However, for fiducial models 
with $\ts=0$, excursions in $\nt$ occur at $\ts=0$, thereby yielding a
zero derivative.  Hence, in this limit, $\nt$ is not a physically 
meaningful parameter and our parameter space is effectively reduced to 12
dimensions.

Finally, we allow the scalar normalization to vary and include
an unknown linear bias.  We describe our normalization choice
in \S\,\ref{sec:diffnorm}; this choice can affect individual derivatives but
does not affect marginalized errors.  The linear bias $b$ is defined
by $P_{\rm gal} = b^2 P_{\rm mass}$.  We also quote results for
$\beta\equiv\om^{0.6}/b$ to facilitate comparisons to results
from peculiar velocity data (\cite{Pee80}\ 1980).

We discuss numerical issues involved with constructing derivatives
with respect to these parameters in Appendix \ref{sec:numerics}.
Here we simply note that our results in Table \ref{tab:lcdm}
are stable to 10\% when step-sizes are halved and that we 
have spot-checked elsewhere with similar results.  Stability
improves as parameter degeneracies are removed, either by the
addition of other data sets or by external priors.

\begin{figure}[tb]
\centerline{\epsfxsize=\colwidth\epsffile{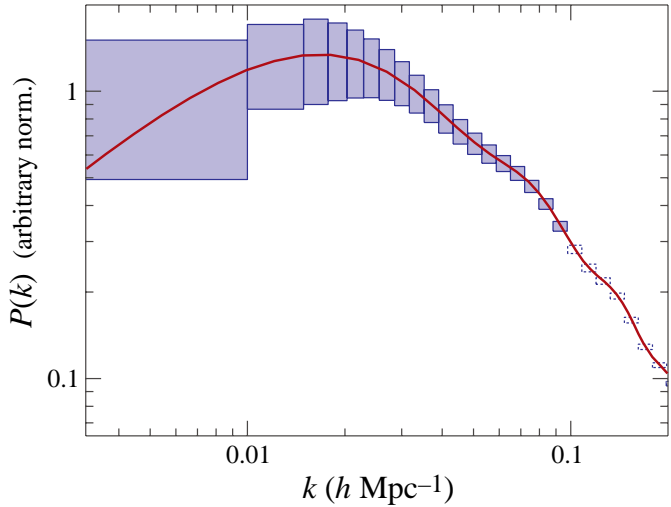}}
\caption{\label{fig:fidsdss}\footnotesize%
The power spectrum for the model of Figure \protect\ref{fig:fidmodel}.
SDSS BRG 1--$\sigma$ error bars are superposed.
Dashed boxes are for $k>\kmax=0.1\ihmpc$; we neglect
information from these scales unless otherwise noted.
}
\end{figure}

\subsection{Fiducial Models}

The Fisher matrix formalism asks how well an experiment can 
distinguish the true (``fiducial'') model of the universe from other models.
The results clearly depend upon the fiducial model itself.  We will
quote results for a number of models, but most of our studies 
are based around a low-density, geometrically flat $\Lambda$CDM
model with $\om=0.35$, $h=0.65$, $\ob=0.05$, $\ol=0.65$, $\on=0.0175$
($m_\nu=0.7\eV$),
$\tau=0.05$, $\ns(\kfid)=1$, and $\ts=0$,
in the region of parameter space favored by recent data.
We display the CMB and matter power spectra for this model in
Figures \ref{fig:fidmodel} and \ref{fig:fidsdss}.
Note that although our fiducial model is flat, we do not restrict our
excursions to flat models: we vary all 12 parameters (including
$\ok$) simultaneously unless explicitly stating otherwise.
Numerical issues are
the reason for the small 
but non-zero values of $\on$ and $\tau$ here (see Appendix 
\ref{sec:numerics}).
We will alter the fiducial values of $\ob$, $\on$,
and $\ts$ in various parameter studies.  In \S\,\ref{sec:othercosms},
we will consider an $\om=1$ SCDM model, a $\om=0.35$ open CDM model, 
and a $\om=0.2$ flat model.  $\alpha=0$ and $\nt=0$ in all models.

\section{CMB alone}\label{sec:cmbalone}

\begin{table*}[th]\footnotesize
\caption{\label{tab:lcdm}}
\begin{center}
{\sc Marginalized Errors for $\Lambda$CDM for various CMB experiments.\\}
\begin{tabular}{lcccc\colskip c\colskip cccc}
\tableskip\tableline\tableline\tableskip
& \multicolumn{4}{c\colskip}{CMB alone} & SDSS & \multicolumn{4}{c}{CMB+SDSS} \\
& \multicolumn{2}{c}{MAP} & \multicolumn{2}{c\colskip}{Planck} & alone
& \multicolumn{2}{c}{MAP} & \multicolumn{2}{c}{Planck}\\
Quantity & Temp & T+P & Temp & T+P & & Temp & T+P & Temp & T+P \\
\tableskip\tableline\tableskip
$h$			 & 1.3 & 0.22 & 1.1 & 0.13 & 1.3 & 0.030 & 0.029 & 0.025 & 0.022 \\
$\om$ 			 & 1.4 & 0.24 & 1.2 & 0.14 & 0.23 & 0.042 & 0.036 & 0.035 & 0.027 \\
$\om h$ 		 & 0.47 & 0.078 & 0.40 & 0.046 & 0.59 & 0.024 & 0.018 & 0.015 & 0.010 \\
$\ol$			 & 1.1 & 0.19 & 0.96 & 0.11 & $\infty$ & 0.056 & 0.042 & 0.036 & 0.024 \\
$\ok$			 & 0.31 & 0.055 & 0.26 & 0.030 & $\infty$ & 0.022 & 0.015 & 0.007 & 0.005 \\[\tskip]
$\ln(\om h^2)$		 & 0.20 & 0.095 & 0.064 & 0.018 & 4.5 & 0.11 & 0.077 & 0.040 & 0.016 \\
$\ln(\ob h^2)$		 & 0.13 & 0.060 & 0.035 & 0.010 & 5.6 & 0.074 & 0.050 & 0.026 & 0.010 \\
$m_\nu$ (eV) $\propto\on h^2$	 & 0.89 & 0.58 & 0.58 & 0.26 & 9.1 & 0.38 & 0.33 & 0.31 & 0.21 \\
$Y_P$			 & 0.020 & 0.020 & 0.018 & 0.013 & \nodata & 0.020 & 0.020 & 0.017 & 0.013 \\[\tskip]
$\ns(\kfid)$ 		 & 0.11 & 0.048 & 0.041 & 0.008 & 1.1 & 0.064 & 0.040 & 0.028 & 0.008 \\
$\ns(H_0)$ 		 & 0.32 & 0.17 & 0.18 & 0.039 & 4.1 & 0.23 & 0.14 & 0.13 & 0.038 \\
$\alpha$		 & 0.030 & 0.018 & 0.015 & 0.004 & 0.42 & 0.022 & 0.015 & 0.011 & 0.004 \\
$\ln\Pinit(\kfid)\equiv\ln\As$		 & 1.4 & 0.43 & 1.1 & 0.073 & $\infty$ & 0.61 & 0.36 & 0.36 & 0.069 \\
$\ln\Pinit(H_0)$		 & 1.8 & 0.71 & 1.3 & 0.16 & $\infty$ & 0.91 & 0.61 & 0.55 & 0.15 \\
$\ts$ 			 & 0.48 & 0.18 & 0.35 & 0.012 & \nodata & 0.24 & 0.16 & 0.15 & 0.012 \\
$\tau$			 & 0.69 & 0.022 & 0.59 & 0.004 & \nodata & 0.27 & 0.021 & 0.21 & 0.004 \\[\tskip]
$\ln\sigma_8$ 		 & 0.48 & 0.14 & 0.42 & 0.057 & $\infty$ & 0.27 & 0.070 & 0.22 & 0.044 \\
$\ln(\sigma_{50}/\sigma_8)$  & 0.86 & 0.15 & 0.75 & 0.093 & 0.27 & 0.029 & 0.028 & 0.024 & 0.020 \\
$\ln \beta$ 		 & \nodata & \nodata & \nodata & \nodata & $\infty$ & 0.27 & 0.068 & 0.20 & 0.027 \\
$\ln b$ 		 & \nodata & \nodata & \nodata & \nodata & $\infty$ & 0.28 & 0.087 & 0.23 & 0.062 \\
\tableskip\tableline
\end{tabular}
\end{center}
NOTES.---%
$\om=0.35$, $\ob=0.05$, $\ol=0.65$, $h=0.65$, $\ts=0$.  
$\nt=0$ and cannot vary.  All errors are $1-\sigma$.
$\kmax=0.1\ihmpc$.  Blank entries indicate that the parameter doesn't
affect the observables for the data set.  
Infinite entries indicate that the parameter affects observables but
is not constrained due to degeneracies; in particular, the growth 
factor, normalization, and bias are all degenerate.
\end{table*}

We begin by presenting the constraints that could in 
principle be achieved by CMB satellites without galaxy
power spectrum information.  Table \ref{tab:lcdm} shows the
results on a variety of quantities using a
$\om=0.35$, $\ol=0.65$ CDM model as our fiducial model.
As expected, CMB data alone provides excellent constraints on 
$\ob h^2$, $\om h^2$, $\ns(\kfid)$ and $\alpha$, as well as on a combination
of $\ok$ and $\ol$.  With polarization information, it also strongly
constrains the reionization optical depth $\tau$.

However, the remaining parameters are nowhere near
``percent level'' in accuracy, even for Planck.
This is because of various degeneracies, particularly the 
angular diameter distance degeneracy 
and the reionization-normalization degeneracy.
Although these were discussed in previous papers, 
none of the tables in 
\cite{Jun96a}\ (1996b), \cite{Zal97a}\ (1997), \cite{Bon97}\ (1997)
or \cite{Cop98}\ (1998) include the case where 
$\ok$ and $\ol$ are measured simultaneously.
As is well known, a combination of changes in $\ol$ and $\ok$
at fixed $\om h^2$ and $\ob h^2$ can keep the angular location 
and morphology of the
acoustic peaks fixed.  This ambiguity causes the value of $h$, $\om$, and
$\om h$ to be poorly constrained (\cite{EHT}); 
moreover, these uncertainties propagate
into the rms density fluctuations $\sigma_R$ 
due to the redshift-space definition of the tophat radius $R$ ($h^{-1}$ Mpc). 
Any additional parameter that affects the angular distance relation,
e.g.\ variations in the equation of state of the missing energy,
creates a similar degeneracy.

Adding polarization information to the CMB helps for most quantities.
Much of the improvement comes from the ability of polarization information
to isolate reionization and tensor contributions (\cite{Zal97a}\ 1997), 
thereby separating
the large-angle temperature effects of $\ol$ and $\ok$.  
Table \ref{tab:lcdm} illustrates that removing this 
ambiguity substantially reduce errors on other quantities
even with the polarization sensitivity of MAP.  

Because the temperature-polarization correlation is not complete,  
polarization also provides independent information on 
the high-redshift, degree-scale
acoustic oscillations to combat sample variance.
This is best illustrated by fixing
$\tau$, $\ts$, and $\ok$, thereby eliminating the need to 
employ the large-angle polarization signal to break degeneracies.  
For the model in Table \ref{tab:lcdm}, 
adding polarization 
information to the temperature data in this restricted 
parameter set improves errors on $\om h^2$,
$\ob h^2$, $\ol$, and $\ns(\kfid)$ by about 40\%-70\% for Planck.
The improvement is only 8\% for MAP, so polarization maps of this
sensitivity are useful only through their constraints on $\tau$
and $\ts$ as discussed above.

Parameter estimation with the CMB alone has been discussed in 
the past in smaller parameter spaces and/or without polarization
information.  As discussed in
Appendix \ref{sec:comparison},
our results disagree with several studies 
in the literature, generally by giving larger error bars.

\section{Complementarity}\label{sec:comp}

\subsection{Adding Redshift Survey Data}\label{sec:cmblss}

The addition of information on the matter power spectrum at the
precision available within the SDSS can make a substantial 
improvement in the error bars on key cosmological quantities.
As seen in Table \ref{tab:lcdm}, one gets large improvements
on $h$, $\om$, and the related quantities $\om h$,  $\Omega_\Lambda$,
and $\Omega_K$, and moderate improvements
in other quantities, particularly $\on h^2$.  Of course, as one
improves the quality of the CMB data set, the fixed level of
SDSS input gets less and less important.

The most striking improvement allowed by measurement of the matter 
power spectrum is the breaking of the angular diameter distance
degeneracy.
Since $\Omega_\Lambda$ and $\Omega_K$
shift the acoustic peaks in opposite directions, to trade one off the other
requires substantial changes in $\Omega_m=1-\Omega_\Lambda-\Omega_K$.  
Because this variation must be done at fixed $\Omega_m h^2$
to maintain the peak morphology, $h$ varies strongly 
in the degenerate direction. A measurement of $h$ thereby breaks
the degeneracy. 

The combination of CMB and galaxy survey data can do just that.
Once $\om h^2$ and $\ob h^2$ are
well-determined by the CMB acoustic peak morphology, 
the real-space power spectrum is 
known.  Varying $h$ causes this pattern to move in redshift space.
Hence, the more features that are present in the matter power
spectrum, the more accurately one can measure $h$ and so 
break the degeneracy.

Within the types of CDM models considered here, the best source
of power spectrum features are the oscillations impressed by
baryonic oscillations.  As shown in \cite{EHT} and \S\,\ref{sec:baryons},
even a 10\% baryon fraction causes large enough features for
SDSS to clamp down on $h$ and $\om$.  Indeed, 
the resulting error bars are nearly as good as if the universe
were {\it assumed} flat in many cases, 
thereby breaking the degeneracy by fiat.

In other sectors, the gains are more modest.  Improvements 
approaching a factor of 2 are possible in $\om h^2$ and $\ns(\kfid)$,
particularly for MAP without polarization.  Polarization tends to
allow the CMB to dominate the constraints, but this does depend on
the value of $\kmax$ and on being able to extract low-$\ell$
cosmological information.  An important complementary aspect
of the data sets enables the determination of the mass of cosmological neutrinos
(\cite{HET}).  We will discuss this further in \S\,\ref{sec:neutrinos}.

As shown in Table \ref{tab:lcdm}, SDSS power spectrum information alone
flounders in this large parameter space.  Details of the shape do 
depend on cosmology, but the features are not well enough detected
for this low baryon fraction and small $\kmax$ (see Figure \ref{fig:fidsdss}).

In all cases, the improvements depend somewhat on the value of $\kmax$,
which we will vary in \S\,\ref{sec:kmax}.
Reverting from the deeper BRG survey to the SDSS main survey degrades the
performance on $h$ and related quantities by about 30\%.

\subsection{The Role of Baryons}\label{sec:baryons}

\begin{table*}[th]\footnotesize
\caption{\label{tab:baryons}}
\begin{center}
{\sc Marginalized Errors as Function of $\ob$.\\}
\begin{tabular}{lcc\colskip cc\colskip cc\colskip cc}
\tableskip\tableline\tableline\tableskip
& \multicolumn{2}{c\colskip}{$\ob=0.005$}
& \multicolumn{2}{c\colskip}{$\ob=0.02$}
& \multicolumn{2}{c\colskip}{$\ob=0.05$}
& \multicolumn{2}{c\colskip}{$\ob=0.10$} \\
Quantity & MAP & +SDSS & MAP & +SDSS & MAP & +SDSS & MAP & +SDSS \\ 
\tableskip\tableline\tableskip
$h$			 & 0.34 & 0.12 & 0.27 & 0.091 & 0.22 & 0.029 & 0.23 & 0.013 \\
$\om$ 			 & 0.36 & 0.12 & 0.29 & 0.086 & 0.24 & 0.036 & 0.25 & 0.018 \\
$\om h$ 		 & 0.12 & 0.039 & 0.093 & 0.029 & 0.078 & 0.018 & 0.084 & 0.009 \\
$\ol$			 & 0.28 & 0.098 & 0.23 & 0.070 & 0.19 & 0.042 & 0.20 & 0.022 \\
$\ok$			 & 0.082 & 0.029 & 0.065 & 0.027 & 0.055 & 0.015 & 0.056 & 0.008 \\[\tskip]
$\ln(\om h^2)$		 & 0.091 & 0.089 & 0.11 & 0.101 & 0.095 & 0.077 & 0.073 & 0.034 \\
$\ln(\ob h^2)$		 & 0.068 & 0.057 & 0.051 & 0.046 & 0.060 & 0.050 & 0.062 & 0.034 \\
$m_\nu$ (eV) $\propto\on h^2$	 & 0.60 & 0.55 & 0.60 & 0.46 & 0.58 & 0.33 & 0.74 & 0.23 \\[\tskip]
$\ns(\kfid)$ 		 & 0.046 & 0.042 & 0.031 & 0.030 & 0.048 & 0.040 & 0.055 & 0.027 \\
$\alpha$		 & 0.032 & 0.020 & 0.023 & 0.017 & 0.018 & 0.015 & 0.026 & 0.015 \\
$\ln\Pinit(\kfid)\equiv\ln\As$		 & 0.39 & 0.35 & 0.30 & 0.28 & 0.43 & 0.36 & 0.47 & 0.23 \\
$\ts$ 			 & 0.11 & 0.11 & 0.15 & 0.14 & 0.18 & 0.16 & 0.17 & 0.13 \\
$\tau$			 & 0.036 & 0.033 & 0.026 & 0.024 & 0.022 & 0.021 & 0.022 & 0.020 \\[\tskip]
$\ln\sigma_8$ 		 & 0.19 & 0.17 & 0.20 & 0.14 & 0.14 & 0.070 & 0.14 & 0.046 \\
$\ln(\sigma_{50}/\sigma_8)$  & 0.27 & 0.031 & 0.19 & 0.026 & 0.15 & 0.028 & 0.13 & 0.026 \\
$\ln \beta$ 		 & \nodata & 0.074 & \nodata & 0.069 & \nodata & 0.068 & \nodata & 0.050 \\
$\ln b$ 		 & \nodata & 0.31 & \nodata & 0.23 & \nodata & 0.087 & \nodata & 0.046 \\
\tableskip\tableline
\end{tabular}
\end{center}
NOTES.---%
All models have $\om=0.35$, $\ol=0.65$, $h=0.65$, and $\ts=0$.  
$\nt=0$ and cannot vary.  All errors are $1-\sigma$.  
CMB data is for MAP with temperature and polarization information.
SDSS column uses information up to $\kmax=0.1\ihmpc$ as well as CMB data.
\end{table*}

Table \ref{tab:baryons} shows the results as a function of the baryon 
fraction.  As the baryon fraction increases, the addition of SDSS 
information becomes more and more helpful.  
$h$ and the related quantities $\om$, $\ol$, $\ok$, and $\sigma_8$ are the most affected,
but even traditional CMB quantities such as $\om h^2$ or $\ob h^2$
see marked improvement at high $\ob/\om$.  

The driving physical effect behind these gains is the structure that
develops in the matter power spectrum due to the high-redshift acoustic
oscillations imprinted by a non-negligible baryon fraction 
(\cite{Pee70} 1970; \cite{Sun70} 1970; 
\cite{Hol89}\ 1989; \cite{Hu96}\ 1996; \cite{Eis98a}\ 1998a).  
In CDM
cosmologies, the baryons and photons oscillate on sub-horizon scales
prior to recombination, while the CDM perturbations simply grow.  The
event of recombination catches the oscillations at various phases
and creates the acoustic peaks we see in the CMB spectrum.  
The perturbations in the baryons
also share this oscillatory history, but in trace-baryon cosmologies it
is erased as the baryons fall into the more-evolved CDM perturbations.
When the baryon fraction is non-negligible, however, the equilibration
of the baryon and CDM perturbations is not completely one-sided, and
the final power spectrum $P(k)$ retains an imprint of the acoustic
oscillations.  The resulting morphology consists of 
a sharp break in the power
spectrum followed by a damped series of wiggles; the whole pattern has a
characteristic scale, known as the sound horizon, which is the distance
a sound wave could travel prior to recombination.

As discussed in \cite{EHT}, once CMB data yields $\omhh$ and $\obhh$,
the physical size of the sound horizon is known.  Measuring this scale
in a redshift survey thereby allows a comparison that gives the 
Hubble constant.  Once the baryon fraction is about 10\% or greater,
SDSS can detect the baryonic features and thereby collapse the
error bars on $h$ and related quantities.

In Table \ref{tab:baryons}, the case of negligible baryons $\ob=0.005$
allows SDSS to make modest improvements over MAP alone.  The factor of
3 improvement on $h$ and $\om$ comes from the two scales that remain
as $\ob$ gets small: the scale of the horizon at matter-radiation 
equality, which is proportional to $\om h$, and the scale of the
horizon when the massive neutrinos become non-relativistic.  The
latter is a result of using a 5\% neutrino fraction; the former
is always present but is inaccurately measured due to confusion 
with spectral tilt.

Baryon fractions exceeding 10\% are strongly favored by observations
of cluster X-ray gas (\cite{Whi93a}\ 1993a; \cite{Dav95}\ 1995; 
\cite{Whi95}\ 1995; \cite{Evr97}\ 1997).  
Lyman $\alpha$ forest theories also favor
high baryon densities (\cite{Wei97}\ 1997); 
when combined with the general observational
preference for low matter densities 
(e.g.\ \cite{Bah97}\ 1997; \cite{Car97a} 1997ab),
this yields a high baryon fraction.
Hence, it seems likely that baryonic features will be prominent
enough in the matter power spectrum that SDSS and perhaps 2dF will 
be able to detect them (\cite{Teg97a}\ 1997a; \cite{Gol98}\ 1998).

If the baryon fraction is yet higher, perhaps 20\%, then the 
detailed morphology of the baryon features can be studied well 
enough to tighten constraints on other, non $h$-related, 
parameters.  The final
columns in Table \ref{tab:baryons} show that a baryon fraction of
$28\%$ could allow SDSS to make a factor of 2 improvement in $\omhh$,
$\obhh$, and $\ns(\kfid)$ over MAP with polarization.

\subsection{Variations in $\om$ and $\ok$}\label{sec:othercosms}

\begin{table*}[th]\footnotesize
\caption{\label{tab:othercosms}}
\begin{center}
{\sc Marginalized Errors for Various Cosmological Models.\\}
\begin{tabular}{lcccc\colskip cccc\colskip cccc}
\tableskip\tableline\tableline\tableskip
& \multicolumn{4}{c\colskip}{SCDM ($\ob=0.1$)}
& \multicolumn{4}{c\colskip}{OCDM ($\om=0.35$)}
& \multicolumn{4}{c}{$\Lambda$CDM ($\om=0.2$)} \\
& \multicolumn{2}{c}{CMB alone} & \multicolumn{2}{c\colskip}{CMB+SDSS} 
& \multicolumn{2}{c}{CMB alone} & \multicolumn{2}{c\colskip}{CMB+SDSS} 
& \multicolumn{2}{c}{CMB alone} & \multicolumn{2}{c}{CMB+SDSS} \\
Quantity & MAP & Planck & MAP & Planck & MAP & Planck & 
MAP & Planck & MAP & Planck & MAP & Planck \\ 
\tableskip\tableline\tableskip
$h$			 & 0.39 & 0.16 & 0.054 & 0.036 & 0.24 & 0.11 & 0.038 & 0.020 & 0.30 & 0.14 & 0.028 & 0.023 \\
$\om$ 			 & 1.5 & 0.64 & 0.20 & 0.15 & 0.24 & 0.12 & 0.038 & 0.023 & 0.15 & 0.074 & 0.015 & 0.012 \\
$\om h$ 		 & 0.38 & 0.16 & 0.055 & 0.038 & 0.075 & 0.038 & 0.020 & 0.008 & 0.062 & 0.030 & 0.011 & 0.005 \\
$\ol$			 & 1.3 & 0.55 & 0.18 & 0.13 & 0.49 & 0.24 & 0.11 & 0.051 & 0.12 & 0.055 & 0.021 & 0.010 \\
$\ok$			 & 0.22 & 0.090 & 0.046 & 0.020 & 0.25 & 0.13 & 0.072 & 0.028 & 0.039 & 0.019 & 0.010 & 0.003 \\[\tskip]
$\ln(\om h^2)$		 & 0.082 & 0.014 & 0.076 & 0.013 & 0.14 & 0.010 & 0.103 & 0.010 & 0.097 & 0.020 & 0.078 & 0.015 \\
$\ln(\ob h^2)$		 & 0.044 & 0.009 & 0.043 & 0.008 & 0.101 & 0.008 & 0.053 & 0.007 & 0.062 & 0.011 & 0.050 & 0.011 \\
$m_\nu$ (eV) $\propto\on h^2$	 & 1.4 & 0.35 & 0.68 & 0.31 & 0.82 & 0.15 & 0.38 & 0.14 & 0.59 & 0.25 & 0.22 & 0.15 \\[\tskip]
$\ns(\kfid)$ 		 & 0.042 & 0.008 & 0.037 & 0.007 & 0.075 & 0.006 & 0.038 & 0.006 & 0.047 & 0.009 & 0.038 & 0.008 \\
$\alpha$		 & 0.016 & 0.004 & 0.015 & 0.004 & 0.043 & 0.003 & 0.022 & 0.003 & 0.027 & 0.004 & 0.016 & 0.004 \\
$\ln\Pinit(\kfid)\equiv\ln\As$		 & 0.38 & 0.067 & 0.34 & 0.066 & 0.59 & 0.050 & 0.33 & 0.049 & 0.42 & 0.077 & 0.34 & 0.067 \\
$\ts$ 			 & 0.24 & 0.019 & 0.22 & 0.019 & 0.29 & 0.016 & 0.25 & 0.016 & 0.13 & 0.009 & 0.12 & 0.009 \\
$\tau$			 & 0.020 & 0.004 & 0.020 & 0.004 & 0.042 & 0.005 & 0.041 & 0.005 & 0.025 & 0.005 & 0.022 & 0.004 \\[\tskip]
$\ln\sigma_8$ 		 & 0.34 & 0.14 & 0.101 & 0.050 & 0.31 & 0.052 & 0.090 & 0.034 & 0.17 & 0.061 & 0.052 & 0.036 \\
$\ln(\sigma_{50}/\sigma_8)$  & 0.42 & 0.17 & 0.042 & 0.034 & 0.23 & 0.071 & 0.037 & 0.016 & 0.13 & 0.076 & 0.027 & 0.021 \\
$\ln \beta$ 		 & \nodata & \nodata & 0.051 & 0.031 & \nodata & \nodata & 0.074 & 0.025 & \nodata & \nodata & 0.076 & 0.034 \\
$\ln b$ 		 & \nodata & \nodata & 0.18 & 0.12 & \nodata & \nodata & 0.11 & 0.052 & \nodata & \nodata & 0.063 & 0.047 \\
\tableskip\tableline
\end{tabular}
\end{center}
NOTES.---%
SCDM model has $\om=1$, $\ob=0.10$, $\ol=0$, $h=0.5$, and $\ts=0$.  
OCDM model has $\om=0.35$, $\ob=0.05$, $\ol=0$, $h=0.65$, $\ns=1.25$, 
and $\ts=0$.  
$\Lambda$CDM model has $\om=0.2$, $\ob=0.03$, $\ol=0.8$, $h=0.8$, and $\ts=0$.
$\nt=0$ and cannot vary.  All errors are $1-\sigma$.  $\kmax=0.1\ihmpc$.
CMB experiments include temperature and polarization information.
\end{table*}

We show the marginalized errors for three other choices of $\om$ and $\ok$
in Table \ref{tab:othercosms}. 
The results are similar to those of the
previous sections, with some
expected variations.  Performance with SDSS on $h$ and $\om$ is
worse (better) in the $\om=1$ SCDM ($\om=0.2$
$\Lambda$CDM) model because the baryonic oscillations in the matter
power spectrum get horizontally shifted to less 
(more) favorable locations relative
to the fixed cutoff scale of $\kmax=0.1\ihmpc$.  On the other
hand, the degree of non-linearity at that scale is not fixed.
Abundances of rich clusters require that low-$\om$ cosmologies
have a higher $\sigma_8$ than high-$\om$ cosmologies
(\cite{Whi93b}\ 1993b; \cite{Via96}\ 1996; \cite{Eke97}\ 1997; 
\cite{Pen97}\ 1997); 
this suggests that a given level of non-linearity should occur at high
values of $k$ in higher $\om$ cases.  In other words, had we fixed
$\kmax$ by requiring that cluster-normalized fluctuations reach a
particular amplitude on that scale, the number of baryon oscillations
in the linear regime, and hence the performance of SDSS, would have
remained more constant.

As expected, the $\om=0.35$ open cosmology does show somewhat 
worse performance than its flat cousin for MAP data, presumably
because the acoustic peaks have been shifted to smaller scales, 
leaving less structure to be resolved by the beam.
This effect is smaller for Planck---while fewer peaks are detected,
the sample-variance errors on scales around the sound horizon are improved.

\subsection{Massive Neutrinos}\label{sec:neutrinos}

\begin{table*}[th]\footnotesize
\caption{\label{tab:neutrinos}}
\begin{center}
{\sc Marginalized Errors as a Function of $\on$\\}
\begin{tabular}{lccc\colskip ccc\colskip ccc\colskip ccc}
\tableskip\tableline\tableline\tableskip
& \multicolumn{3}{c\colskip}{$\on=0.01$} 
& \multicolumn{3}{c\colskip}{$\on=0.05$} 
& \multicolumn{3}{c\colskip}{$\on=0.10$} 
& \multicolumn{3}{c\colskip}{$\on=0.20$} \\
& \multicolumn{3}{c\colskip}{($m_\nu=0.24\eV$)} 
& \multicolumn{3}{c\colskip}{($m_\nu=1.2\eV$)} 
& \multicolumn{3}{c\colskip}{($m_\nu=2.4\eV$)} 
& \multicolumn{3}{c\colskip}{($m_\nu=4.7\eV$)} \\
& CMB & \multicolumn{2}{c\colskip }{+SDSS} 
& CMB & \multicolumn{2}{c\colskip }{+SDSS} 
& CMB & \multicolumn{2}{c\colskip }{+SDSS} 
& CMB & \multicolumn{2}{c\colskip }{+SDSS} \\
Quantity & T+P & 0.1 & 0.2 & T+P & 0.1 & 0.2 
& T+P & 0.1 & 0.2 & T+P & 0.1 & 0.2 \\
\tableskip\tableline\tableskip
$m_\nu$ (eV) $\propto\on h^2$	 & 1.1 & 0.81 & 0.42 & 1.4 & 0.68 & 0.30 & 5.5 & 0.84 & 0.31 & 6.8 & 1.7 & 0.38 \\
$\ln(\om h^2)$		 & 0.11 & 0.101 & 0.082 & 0.082 & 0.076 & 0.063 & 0.067 & 0.065 & 0.053 & 0.084 & 0.059 & 0.044 \\[\tskip]
$h$			 & 0.38 & 0.065 & 0.020 & 0.39 & 0.054 & 0.018 & 0.39 & 0.046 & 0.017 & 0.38 & 0.039 & 0.015 \\
$\om$ 			 & 1.5 & 0.28 & 0.092 & 1.5 & 0.20 & 0.068 & 1.5 & 0.16 & 0.055 & 1.5 & 0.12 & 0.039 \\
\tableskip\tableline
  \tableskip
$m_\nu$ (eV) $\propto\on h^2$	 & 0.53 & 0.47 & 0.29 & 0.35 & 0.31 & 0.19 & 0.30 & 0.28 & 0.17 & 0.31 & 0.31 & 0.17 \\
$\ln(\om h^2)$		 & 0.028 & 0.025 & 0.018 & 0.014 & 0.013 & 0.011 & 0.010 & 0.010 & 0.010 & 0.009 & 0.009 & 0.009 \\[\tskip]
$h$			 & 0.16 & 0.048 & 0.016 & 0.16 & 0.036 & 0.014 & 0.16 & 0.028 & 0.012 & 0.16 & 0.021 & 0.009 \\
$\om$ 			 & 0.64 & 0.21 & 0.075 & 0.64 & 0.15 & 0.058 & 0.64 & 0.11 & 0.048 & 0.64 & 0.081 & 0.034 \\
  \tableskip\tableline
\end{tabular}
\end{center}
NOTES.---%
$\om=1$, $\ob=0.1$, $\ol=0$, $h=0.5$, and $\ts=0$ for all columns. 
We assume one species of massive neutrino.
$\nt=0$ and does not vary.  All errors are $1-\sigma$.  
All columns include CMB information on temperature and polarization;
the top four lines are for MAP, the bottom four for Planck.
SDSS columns include information to $\kmax=0.1\ihmpc$ or
$\kmax=0.2\ihmpc$, as noted.
\end{table*}

\begin{figure}[tb]
\centerline{\epsfxsize=\colwidth\epsffile{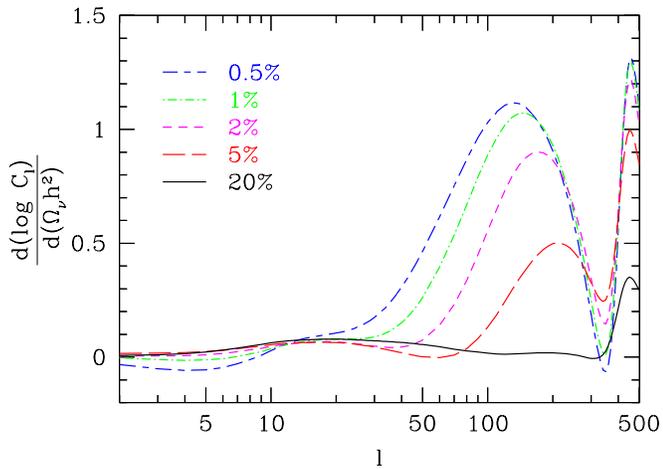}}
\caption{\label{fig:neutrino}\footnotesize%
The derivative $d(\ln C_{T\ell})/d\on h^2$ as a function of 
neutrino fraction $\on/\om$.  Top to bottom (for the first peak):
0.5\%, 1\%, 2\%, 5\%, and 20\%.  The cosmology is the $\om=0.35$
$\Lambda$CDM model, but with $\tau=0.1$, for a variety of $\on$.}
\end{figure}

Massive neutrinos present particular problems for Fisher matrix
analyses of these data sets.
Varying the neutrino mass changes the free-streaming scale,
below which 
their rms velocity prevents them from clustering. 
As $\on$ (or equivalently $m_\nu$) 
approaches zero, the neutrinos become
fully relativistic and the free-streaming 
scale of the neutrinos approaches the horizon scale
(\cite{Bon83} 1983; \cite{Hol89} 1989; \cite{Ma95}\ 1995; \cite{Dod96}\ 1996; 
\cite{Hu98a}\ 1998).  
This means that at large scales, it takes only a small upward variation
of the neutrino mass from zero to bring the neutrinos out of the
free-streaming regime. 
As the power spectra phenomenology differs between these
two physical regimes, we expect that the
derivative of $P(k)$ or $C_l$ with respect to $\Omega_\nu h^2$ will be
sharply different above and below this small mass threshold, which is
itself a function of scale.

This raises both a practical and a theoretical concern.  First, given
numerical noise, is it possible to calculate the derivative as
$\Omega_\nu$ goes to zero, or must one always difference pairs of
models that bracket the mass transition for some scale of interest?
Second, since the derivative is changing significantly even for small
values of $\on h^2$, how should we interpret the resulting error
bars on $\on h^2$?  The difficulty is that 
the log-likelihood function is not well-approximated
by a quadratic expansion around $\on=0$
relative to the quality of the constraints. 

This situation is shown in Figure \ref{fig:neutrino}.  The temperature
$C_\ell$ derivative with respect to the massive neutrino density is
a strong function of the fiducial value of $\on$.  
For small $\on$, the derivative has a
large peak at $\ell\approx150$, but for larger $\on$, the feature goes
away entirely.  This means that if one began at $\on=0$ and increased
the neutrino mass, the power at $\ell=150$ would change rapidly at
first and then saturate.  A Fisher matrix analysis around $\on\approx0$
would show strong limits on $\on$, while an analysis around
$\on/\om\approx0.2$ would show weak limits.

\begin{table*}[th]\footnotesize
\caption{\label{tab:tensors}}
\begin{center}
{\sc Marginalized Errors as a function of Tensor-to-Scalar Ratio\\}
\begin{tabular}{lcc\colskip ccc\colskip c\colskip c}
\tableskip\tableline\tableline\tableskip
& \multicolumn{5}{c\colskip}{$\ts=1.0$} & $\ts=0.3$ & $\ts=0.1$ \\
& \multicolumn{2}{c\colskip }{MAP} & \multicolumn{3}{c\colskip}{Planck}
& Planck & Planck \\
Quantity & T+P & SDSS & no B & T+P & SDSS & SDSS & SDSS \\
\tableskip\tableline\tableskip
$\ts$ 			 & 2.0 & 0.79 & 0.45 & 0.42 & 0.28 & 0.13 & 0.064 \\
$X=\ts+x\nt$ & 0.65 & 0.54 & 0.20 & 0.16 & 0.14 & 0.055 & 0.032 \\
$\nt$			 & 0.40 & 0.16 & 0.12 & 0.100 & 0.080 & 0.14 & 0.24 \\[\tskip]
$\ns(\kfid)$ 		 & 0.065 & 0.047 & 0.012 & 0.012 & 0.011 & 0.009 & 0.008 \\
$\ns(H_0)$ 		 & 0.45 & 0.21 & 0.059 & 0.055 & 0.051 & 0.043 & 0.040 \\
$\alpha$		 & 0.048 & 0.022 & 0.005 & 0.005 & 0.005 & 0.004 & 0.004 \\
$\ln\Pinit(\kfid)\equiv\ln\As$		 & 0.59 & 0.43 & 0.11 & 0.103 & 0.095 & 0.080 & 0.074 \\
$\ln\Pinit(H_0)$		 & 1.4 & 0.81 & 0.27 & 0.25 & 0.23 & 0.18 & 0.17 \\
\tableskip\tableline\tableskip
Value of $x$ & 4.9 & 3.6 & 3.4 & 3.7 & 3.1 & 0.85 & 0.23 \\
\tableskip\tableline
\end{tabular}
\end{center}
NOTES.---%
$\om=0.35$, $\ob=0.05$, $\ol=0.65$, $h=0.65$, $\nt=0$.
All errors are $1-\sigma$.  $\kmax=0.1\ihmpc$.
SDSS columns include CMB data with polarization including B channel.
The ``No B'' column presents results in which B-channel polarization 
has been ignored.  The B channel is of negligible importance for MAP.
$x$ has been chosen so as to minimize the error on the quantity $X$;
for this choice, $X$ is uncorrelated with $\nt$ and the error is
equal to the error on $\ts$ if $\nt$ were held fixed.
Without CMB polarization information, errors on $\ts$ from the MAP or
Planck are much larger, but errors when combined with SDSS are similar to those
in the MAP+SDSS column. 
\end{table*}

The derivative of the matter power spectrum with respect to the 
neutrino mass shows a similar behavior: it is non-zero only below
a break scale that shifts to small $k$ as $\on h^2$ gets smaller.
Because of degeneracies, the break itself needs to be detected.
Redshift-survey data is less restrictive at small $k$.
Hence, whereas CMB data is more restrictive at low $\on$, redshift
survey data is more restrictive at high $\on$.

We are interested in what neutrino masses can be distinguished from
zero.  Because of the above trend, even if a data set preferred $\on=0$,
the upper limit on the neutrino mass would be underestimated.  To be
conservative, we
run at $\on\ne0$ instead; the significance at which $\on=0$
can be excluded by the data set is then underestimated.

In Table \ref{tab:neutrinos}, we present marginalized errors as a 
function of $\on$ within a $\om=1$ CDM model.  MAP data alone 
does not detect the mass of the neutrino in any of the cases.
In particular, as $\on$ increases, the angular scale at which the
neutrinos have an observable signature drops below the resolution
of the experiment.  Planck can continue to track the effect and
therefore can detect the neutrino mass.  In both data sets, as
$\on$ increases, $\om h^2$ errors tend to decrease and errors on
the tilt sector remain unchanged.

With SDSS data added to MAP, neutrino masses exceeding
$\sim\!1\eV$ would be detectable (\cite{HET}).  SDSS also improves
the limits of Planck by up to a factor of 2.  Sensitivity to lower
masses is possible for $\om<1$.  Unlike the situation
with the $\Lambda$CDM models presented elsewhere in this paper,
$\kmax\approx0.2\ihmpc$ may be appropriate for the neutrino signatures
of SCDM: these models have lower normalization (e.g.\ cluster abundances
suggest $\sigma_8\approx0.5$ not 1.0)
and thus less non-linear
contamination. Also, because the neutrino signature is not oscillatory, 
it may be harder for nonlinear dynamics to wash it out.

As $\on$ increases, performance on $\om$ and $h$ from the combination
of SDSS and CMB also improves.  Like the baryons, massive neutrinos
impress a scale on the matter power spectrum: there is a break in the
spectral index around the scale of the horizon at the epoch when the
neutrinos become non-relativistic.  This physical scale depends upon
$\om h^2$ and $\on/\om$; once these two are determined, one gains
leverage on $h$ and $\om$ by detecting the scale in redshift space.

We have run $\on/\om=0.05$ unless otherwise specified.  With CMB and
redshift survey data, this model is distinguished from $\Omega_\nu=0$
at $2-4\sigma$.  Larger neutrino fractions rely more and more on 
the matter power spectrum data to provide constraints.

\subsection{Tensors}\label{sec:tensors}

Interpreting the errors on the tensor-to-scalar ratio $T/S$ requires
special care since we have at present no guide as to its value in the
real universe.  Although in power-law inflation $T/S \approx 7(1-\ns)$
(\cite{Dav92}\ 1992), many inflationary models predict $T/S \approx 0$
(\cite{Lyt97}\ 1997) and in general slow-roll inflationary models obey
$T/S \approx -7\nt$ (see e.g.~\cite{Lid93}\ 1993).   For this reason,
we conduct a parameter study of $T/S$ here.

{\it $T/S$ Sequence---}As $\ts$ increases in the fiducial model, 
the errors on most quantities
degrade slightly due to the reduction of small-angle anisotropy signal
at fixed COBE-normalization.  With information from SDSS, the 
degradation is smaller yet.  However, the errors on $\ts$ and $\nt$
are strong functions of $\ts$.  Errors on $\nt$ and {\it fractional}
errors on $\ts$ decrease as $\ts$ increases,
the latter of course becoming infinite as $\ts\to 0$.
Including SDSS information
significantly reduces the error bars even in the case of Planck; 
for example, better controlling the value of $\ol$ and $\ok$ substantially
helps Planck to sort out the various large-angle signals and approach
the fixed cosmology limit of \cite{Kno95} (1995) 
(see also \cite{Kin98}\ 1998).  
We show results for $\ts=1$, 0.3, and 0.1 in Table \ref{tab:tensors}.
Remember, however, that each of these assumes $\nt=0$ in the 
fiducial model; the inflationary consistency relation 
predicts $\nt<0$ for $\ts\ne0$, in which case the signal would be slightly
overestimated.  

{\it $T/S$--$\nt$ plane---}In 
the presence of a varying $\nt$, the value and hence the error
bars on the tensor-to-scalar ratio become scale-dependent.  The
ratio of quadrupole power $\ts$ is a pessimistic choice because the
ability of the data sets to constrain tensor contributions comes
at somewhat smaller angles ($\ell\approx50$, see also
\cite{Whi96}\ 1996).  Hence, variations in
$\nt$ and $\ts$ are highly anti-correlated, and the error bars on $\ts$
do not properly reflect the ability of the experiments to detect
the tensor signal.  Other choices of parameters, e.g. the ratio
of tensor power to scalar power at $\ell=50$, should yield smaller errors.
Appendix \ref{sec:minerror} discusses this situation.
One can write the general combination of $\nt$ and $\ts$
as $X\equiv \ts + x\nt$.  One then varies $x$ to minimize the errors
on $X$.  At this minimum, the uncertainties on $X$ are uncorrelated
with those on $\nt$; moreover, the marginalized errors on $X$ (with 
$\nt$ varying) are
identical to those on $\ts$ in the case where $\nt$ is held fixed.
The errors on the best choice of $X$ are shown in Table \ref{tab:tensors}
and better reflect the constraint on the tensor signal.

{\it Detection Threshold---}
For this set of $\Lambda$CDM
models, the tensor signal could be detectable with Planck
at high significance for $\ts$ as low as $\sim\!0.1$
(see Tab.\ \ref{tab:tensors}).
Only strong tensor signals ($\ts\gtrsim1$) can be isolated by MAP. 
Redshift-survey information can help, especially in case the 
tensor polarization signal is obscured by foregrounds.
Achieving the best performance depends on being able to separate 
the $E$ and $B$
channels of polarization (\cite{Zal97a}\ 1997; \cite{Kos98} 1998); 
for example, if the $B$-channel is
completely ignored, then the $X$ errors on the $\ts=0.1$ model
in Table \ref{tab:tensors} increase by 50\%.  However, much of the
leverage comes from temperature data; if the uncertainty from 
other large-angle contributors can be removed, the tensor plateau 
at $\ell\approx50$ can be detected.  
For example, as the optical depth in the fiducial model increases,
the detection threshold for $T/S$ drops because $\tau$ can be better
constrained from the temperature data itself (\cite{Zal97a}\ 1997).

{\it Consistency Relation---}If the tensor-to-scalar ratio approaches unity, 
the tensor tilt can be measured as well (\cite{Kno95} 1995;
\cite{Zal97a}\ 1997).  
This could allow a test of the inflationary 
prediction $T/S \approx -7 \nt$.  
In attempting to test specific inflationary models,
e.g.\ the $\ts\approx7(1-\ns)$ relation of power-law inflation,
one should estimate $\ts$ and $\ns$ at the same scale.
If $\alpha\ne0$, the scalar tilt at scales somewhat larger 
than $\kfid$ will be more uncertain that $\ns(\kfid)$,
leading to weaker constraints on inflation.

\section{Assumptions}\label{sec:assumptions}

\subsection{CMB Foregrounds}\label{sec:foregrounds}

We have assumed that a number of the frequency channels measured
by the CMB experiments will be used for foreground removal
(e.g., \cite{Bra94}\ 1994; \cite{TE96}\ 1996; \cite{Ber96}\ 1996;
\cite{Teg98b}\ 1998; \cite{Hob98}\ 1998),
leaving only the subset listed in Table \ref{tab:specs} available
for cosmology.
Assuming that foregrounds can be eliminated to this level
may be optimistic, especially for 
the polarization at the largest angular 
scales and for small angular scales in general.  For the former, 
the cosmic signal is small (see Fig.\ \ref{fig:fidmodel} inset)
and must be detected 
against potentially larger Galactic foregrounds (\cite{Kea98}\ 1997).  
Large-angle temperature signals are also 
useful for distinguishing curvature
effects but will in any case be severely limited by 
cosmic variance.  
On the smallest angular scales, point source subtraction may be
insufficient especially where the cosmic signal is falling due to the
finite duration of last scattering 
(\cite{Tof98}\ 1998; \cite{Gui98}\ 1998; 
\cite{TegOli98}\ 1998; \cite{Ref98}\ 1998).

If we eliminate even more frequency channels from the Fisher matrix,
the errors on cosmological parameters for the model in Table \ref{tab:lcdm}
increase only slightly.
Restricting MAP to use only the 90 GHz channel for temperature data
increases errors bars by less than 10\%.  Using only the 90 GHz channel
for both the temperature and the polarization data is a more serious
loss: 50\% on most parameters and 150\% on $\tau$.  Essentially, the 90
GHz channel is nearly sample-variance limited for the temperature
anisotropies, so little is added by the lower frequencies, while the
channels are all roughly equally important for the noise-dominated
large-angle polarization signal.

\begin{table*}[th]\footnotesize
\caption{\label{tab:kmax}}
\begin{center}
{\sc Marginalized Errors as Function of $\kmax$.\\}
\begin{tabular}{lc\colskip cccc\colskip ccc}
\tableskip\tableline\tableline\tableskip
& & \multicolumn{4}{c\colskip}{Linear $P(k)$}
& \multicolumn{3}{c}{Smooth $P(k)$} \\
Quantity & MAP & 0.05 & 0.1 & 0.2 & 0.4 & 0.1 & 0.2 & 0.4 \\
\tableskip\tableline\tableskip
$h$			 & 0.22 & 0.096 & 0.029 & 0.012 & 0.009 & 0.100 & 0.089 & 0.085 \\
$\om$ 			 & 0.24 & 0.098 & 0.036 & 0.016 & 0.014 & 0.10 & 0.100 & 0.099 \\
$\om h$ 		 & 0.078 & 0.033 & 0.018 & 0.011 & 0.009 & 0.036 & 0.036 & 0.036 \\
$\ol$			 & 0.19 & 0.081 & 0.042 & 0.024 & 0.021 & 0.088 & 0.087 & 0.086 \\
$\ok$			 & 0.055 & 0.030 & 0.015 & 0.010 & 0.009 & 0.029 & 0.022 & 0.020 \\[\tskip]
$\ln(\om h^2)$		 & 0.095 & 0.094 & 0.077 & 0.054 & 0.049 & 0.088 & 0.071 & 0.065 \\
$\ln(\ob h^2)$		 & 0.060 & 0.058 & 0.050 & 0.038 & 0.035 & 0.058 & 0.049 & 0.044 \\
$m_\nu$ (eV) $\propto\on h^2$	 & 0.58 & 0.55 & 0.33 & 0.17 & 0.17 & 0.43 & 0.43 & 0.42 \\[\tskip]
$\ns(\kfid)$ 		 & 0.048 & 0.048 & 0.040 & 0.029 & 0.027 & 0.045 & 0.039 & 0.037 \\
$\alpha$		 & 0.018 & 0.018 & 0.015 & 0.012 & 0.007 & 0.018 & 0.013 & 0.008 \\
$\ln\Pinit(\kfid)\equiv\ln\As$		 & 0.43 & 0.43 & 0.36 & 0.26 & 0.24 & 0.41 & 0.35 & 0.33 \\
$\ts$ 			 & 0.18 & 0.18 & 0.16 & 0.14 & 0.14 & 0.17 & 0.17 & 0.17 \\
$\tau$			 & 0.022 & 0.022 & 0.021 & 0.021 & 0.021 & 0.022 & 0.021 & 0.021 \\[\tskip]
$\ln\sigma_8$ 		 & 0.14 & 0.12 & 0.070 & 0.038 & 0.034 & 0.12 & 0.11 & 0.102 \\
$\ln(\sigma_{50}/\sigma_8)$  & 0.15 & 0.066 & 0.028 & 0.011 & 0.010 & 0.030 & 0.012 & 0.011 \\
$\ln \beta$ 		 & \nodata & 0.090 & 0.068 & 0.053 & 0.045 & 0.094 & 0.093 & 0.091 \\
$\ln b$ 		 & \nodata & 0.22 & 0.087 & 0.041 & 0.035 & 0.24 & 0.22 & 0.21 \\
\tableskip\tableline
\end{tabular}
\end{center}
NOTES.---%
All models have $\om=0.35$, $\ob=0.05$, $\ol=0.65$, $h=0.65$, and $\ts=0$.  
$\nt=0$ and cannot vary.  All errors are $1-\sigma$.  
CMB data for MAP with temperature and polarization information included
on all columns.  Linear $P(k)$ means using actual linear power spectrum.
Smooth $P(k)$ columns use a fitting formula that captures the break
at the sound horizon but eliminates the baryon oscillations in $P(k)$.
\end{table*}

For Planck, retaining only the 143 GHz channel for cosmology increases
the error bars by $\sim\!10\%$ for temperature only and by $\sim\!15\%$
for temperature and polarization (but 25\% on $\ts$).  Retaining only
the 217 GHz channel does a little better ($\sim\!7\%$) on temperature
and somewhat worse ($\sim\!25\%$) on polarization (and a factor of 3 on
$\ts$).  Planck's expected polarization performance doesn't quite reach
the sample-variance limit, but a single channel of temperature data
saturates the limit down to a beam scale that varies only
slightly between the prime channels.

Conversely, if one adds the channels reserved for foreground
subtraction back into the cosmological Fisher matrix, the cosmological
error bars don't improve significantly.  Even if the remaining eight
channels on Planck were assumed to give cosmological signal, including
them would reduce the error bars in Table \ref{tab:lcdm} by only
$\sim\!1\%$!  MAP temperature results are similarly insensitive to the
20 GHz and 30 GHz channels, although the polarization signal can be helped
(40\% improvement on $\tau$, less on other quantities).
Reserving these channels for foreground subtractions thus comes 
at little cost for cosmology.

\subsection{Gravitational Lensing}
\label{sec:lensing}
In this paper, we ignore the effects of gravitational lensing on the 
CMB power spectrum.   
Gravitational lensing from large scale structure smooths out 
the high multipoles of the CMB 
(\cite{Bla87} 1987; \cite{Col89} 1989;
\cite{Sel96a}\ 1996a) in a way that is dependent on the amplitude
of the matter power spectrum at low redshift. 
Adjusting $\ol$ and $\ok$ to keep
a constant angular diameter distance allows large changes in
this amplitude
thereby breaking the angular diameter distance
degeneracy (\cite{Met97}\ 1997). 
However, the effect is small and even Planck cannot
use it to attain good estimates on $\om$ and $h$ (\cite{Sto98}\ 1998).
We neglect it here since
the degeneracy is broken much more effectively by 
additional information from redshift surveys or other sources.

\subsection{Varying $\kmax$}\label{sec:kmax}

We have assumed that the galaxy power spectrum follows linear theory
on scales longward of a wavenumber $\kmax$.  Non-linear evolution
on smaller scales may obscure the cosmological information in the
linear power spectrum.  To be conservative, we have neglected all
cosmological information on smaller scales.

Baryonic features in the linear power spectrum are of particular
importance for cosmological parameter estimation, yet they are 
washed out at second-order in perturbation theory (e.g.\ \cite{Jai94}\ 1994).
\cite{EHT} found that fluctuation levels up to 
$k^3 P(k)/2\pi^2\approx0.5$ preserved the features for the model
of Table \ref{tab:lcdm}. 
Simulations give similar results (\cite{Mei98}\ 1998).  
Hence, for this model, with either CMB or cluster
abundance normalization, one may expect linear theory to apply to
$\kmax=0.1\ihmpc$ but not much beyond.  This is enough to see the first
acoustic oscillation.

\begin{table*}[th]\footnotesize
\caption{\label{tab:priors}}
\begin{center}
{\sc Marginalized Errors as Function of Parameter Space.\\}
\begin{tabular}{lccccccc\colskip ccccccc}
\tableskip\tableline\tableline\tableskip
Quantity & \multicolumn{7}{c\colskip}{MAP (TP)}
& \multicolumn{7}{c}{MAP + SDSS} \\
\tableskip\tableline\tableskip
$h$			 & 0.22 & 0.22 & 0.064 & 0.052 & 0.063 & 0.048 & 0.022 & 0.029 & 0.029 & 0.024 & 0.014 & 0.024 & 0.014 & 0.012 \\
$\om$ 			 & 0.24 & 0.23 & 0.098 & 0.081 & 0.097 & 0.074 & 0.034 & 0.036 & 0.034 & 0.036 & 0.020 & 0.036 & 0.019 & 0.016 \\
$\om h$ 		 & 0.078 & 0.076 & 0.042 & 0.035 & 0.041 & 0.031 & 0.015 & 0.018 & 0.015 & 0.015 & 0.008 & 0.015 & 0.008 & 0.007 \\
$\ol$			 & 0.19 & 0.18 & 0.098 & 0.081 & 0.097 & 0.074 & 0.034 & 0.042 & 0.035 & 0.036 & 0.020 & 0.036 & 0.019 & 0.016 \\
$\ok$			 & 0.055 & 0.054 & \nodata & \nodata & \nodata & \nodata & \nodata & 0.015 & 0.012 & \nodata & \nodata & \nodata & \nodata & \nodata \\[\tskip]
$\ln(\om h^2)$		 & 0.095 & 0.069 & 0.086 & 0.073 & 0.084 & 0.065 & 0.035 & 0.077 & 0.059 & 0.033 & 0.020 & 0.033 & 0.018 & 0.018 \\
$\ln(\ob h^2)$		 & 0.060 & 0.042 & 0.055 & 0.054 & 0.053 & 0.050 & 0.025 & 0.050 & 0.035 & 0.028 & 0.028 & 0.028 & 0.028 & 0.023 \\
$m_\nu$ (eV) $\propto\on h^2$	 & 0.58 & 0.53 & 0.58 & \nodata & 0.56 & \nodata & \nodata & 0.33 & 0.30 & 0.31 & \nodata & 0.31 & \nodata & \nodata \\
$Y_P$			 & 0.020 & 0.020 & 0.020 & \nodata & 0.020 & \nodata & \nodata & 0.020 & 0.020 & 0.020 & \nodata & 0.020 & \nodata & \nodata \\[\tskip]
$\ns(\kfid)$ 		 & 0.048 & 0.028 & 0.045 & 0.039 & 0.045 & 0.037 & 0.014 & 0.040 & 0.024 & 0.021 & 0.018 & 0.021 & 0.018 & 0.012 \\
$\ns(H_0)$ 		 & 0.17 & 0.15 & 0.16 & 0.14 & 0.045 & 0.037 & 0.014 & 0.14 & 0.12 & 0.14 & 0.14 & 0.021 & 0.018 & 0.012 \\
$\alpha$		 & 0.018 & 0.018 & 0.016 & 0.016 & \nodata & \nodata & \nodata & 0.015 & 0.014 & 0.014 & 0.014 & \nodata & \nodata & \nodata \\
$\ln\Pinit(\kfid)\equiv\ln\As$		 & 0.43 & 0.25 & 0.40 & 0.35 & 0.40 & 0.34 & 0.12 & 0.36 & 0.21 & 0.18 & 0.15 & 0.18 & 0.15 & 0.101 \\
$\ln\Pinit(H_0)$		 & 0.71 & 0.29 & 0.71 & 0.57 & 0.61 & 0.51 & 0.18 & 0.61 & 0.23 & 0.49 & 0.46 & 0.27 & 0.24 & 0.15 \\
$\ts$ 			 & 0.18 & \nodata & 0.17 & 0.16 & 0.16 & 0.15 & \nodata & 0.16 & \nodata & 0.12 & 0.12 & 0.087 & 0.085 & \nodata \\
$\tau$			 & 0.022 & 0.021 & 0.022 & 0.021 & 0.021 & 0.020 & 0.020 & 0.021 & 0.020 & 0.021 & 0.021 & 0.020 & 0.020 & 0.020 \\[\tskip]
$\ln\sigma_8$ 		 & 0.14 & 0.14 & 0.13 & 0.056 & 0.12 & 0.052 & 0.035 & 0.070 & 0.069 & 0.062 & 0.027 & 0.062 & 0.027 & 0.027 \\
$\ln(\sigma_{50}/\sigma_8)$  & 0.15 & 0.15 & 0.074 & 0.059 & 0.069 & 0.057 & 0.033 & 0.028 & 0.028 & 0.027 & 0.017 & 0.027 & 0.015 & 0.015 \\
$\ln \beta$ 		 & \nodata & \nodata & \nodata & \nodata & \nodata & \nodata & \nodata & 0.068 & 0.058 & 0.047 & 0.046 & 0.045 & 0.043 & 0.038 \\
$\ln b$ 		 & \nodata & \nodata & \nodata & \nodata & \nodata & \nodata & \nodata & 0.087 & 0.087 & 0.082 & 0.031 & 0.080 & 0.031 & 0.029 \\
\tableskip\tableline
\end{tabular}
\end{center}
NOTES.---%
$\om=0.35$, $\ob=0.05$, $\ol=0.65$, $h=0.65$, and $\ts=0$.  
$\nt=0$ and cannot vary.  All errors are $1-\sigma$.  
CMB data for MAP with temperature and polarization information included
on all columns.  SDSS columns use CMB data and $\kmax=0.1\ihmpc$.
Blank entries indicate that the parameter has been held fixed.
\end{table*}

Because the appropriate value of $\kmax$ depends on normalization,
we show the results for our fiducial $\Lambda$CDM model as a
function of $\kmax$ in Table \ref{tab:kmax}.  
As expected, the results
do depend strongly on $\kmax$.  In particular, as $\kmax$ increases
from $0.05\ihmpc$ to $0.2\ihmpc$, the errors on $h$ and related quantities
drop sharply.  In this model, the break in the power spectrum occurs
at about $0.03\ihmpc$ and the first peak is at $0.08\ihmpc$ 
(Fig.\ \ref{fig:fidsdss}).
Little information is added between 
$\kmax=0.2\ihmpc$ and $\kmax=0.4\ihmpc$ here.  On these small scales,
the acoustic oscillations have been damped away even in linear theory.

While our choice of $\kmax$ may be appropriate for baryonic features,
other aspects of cosmology may not be so fragile.  For example, the
broadband level of the linear power spectrum, affected by $\ns$
and $\alpha$, could potentially be reconstructed from the quasi-linear
regime (\cite{Pea94}\ 1994).
As a means of exploring this, we consider an alteration to our
treatment of the matter power spectrum, replacing the true linear power
spectrum by a smoothed spectrum based on the fitting formula of
\cite{Eis98b}\ (1998b).  This formula preserves the sharp break at the sound
horizon ($k\approx0.03\ihmpc$ here) but ignores all the smaller-scale
oscillations.  The resulting marginalized errors are considerably
worse, typically equivalent to using the true power spectrum out to 
$\kmax\approx0.05\ihmpc$.  Moreover, the results show little improvement
as $\kmax$ increases from $0.1\ihmpc$ to $0.4\ihmpc$.  
In other words, within the constraints available through CMB satellites, 
little additional cosmological information is gained by detection
of a featureless matter power spectrum.  Only detection of the
acoustic oscillations or spectral breaks (e.g.\ massive neutrinos)
produces significant improvements.  An exception to this conclusion
is the error bar on $\alpha$, the running of the scalar tilt, which
continues to improve as $\kmax$ increases.

\subsection{Smaller Parameter Spaces}\label{sec:priors}

We next consider the effects of removing parameters from our model space.
Removing parameters in the face of degeneracies is often motivated by
Occam's razor or on the grounds that external information will 
eliminate some of the options available.  In addition, doing so
allows us to explore the extent of the degeneracies and facilitates 
comparison to previous works.

Removing parameters does not require recalculation of models; 
rather, it means that we hold their value fixed at the fiducial
value (by a prior) 
while the errors on other quantities are computed.  In Table
\ref{tab:priors}, we show several different cases, where
parameters held fixed are denoted by blank entries,
both with and
without redshift survey information.  The error bars necessarily
decrease as degrees of freedom are removed.  It is important
to remember that improvements found by removing a parameter depend
on which the remaining parameters are; in other words, statements such as
``removing $\alpha$ is negligible'' apply only to the particular
context of our 12-dimensional parameter space.

{\it Tensors}---Eliminating tensors from the model (i.e.~assuming $\ts=0$) 
allows $\ns$, $\omhh$, and $\obhh$ to be better determined by MAP.
Planck's longer lever arm allows it to constrain $\ns$ regardless of
tensors.

{\it Curvature}---Fixing the curvature of course breaks the angular
diameter distance degeneracy by assumption.  Therefore, CMB data alone
can turn the location of the acoustic peaks into a measure of $h$,
$\om$, and other quantities.  It is interesting to note, however, that
MAP doesn't get enough accuracy on, e.g., $\omhh$ to keep uncertainties
in the sound horizon from propagating into the measure of the angular
diameter distance.  Planck without polarization does only slightly
better in this regard.

If SDSS information is included, assuming $\ok=0$ makes
very little difference to errors on $h$ and $\om$,
since the angular distance degeneracy is already broken.  
It does however affect other parameters that 
determine the peak locations such as $\obhh$, 
$\omhh$ and $\ns(\kfid)$. To see why this is so, consider
the general case where $\ok$ is allowed to vary.
One has two uncertain quantities, $\ol$
and $\ok$, to map a fairly well-constrained quantity, the
sound horizon, to the location of the peaks in $C_\ell$ and $P(k)$.
This yields good constraints on $\ol$ and $\ok$.  Now, with $\ok=0$
assumed, these two observations can both constrain $\ol$ {\it and}
reduce the remaining uncertainties on the sound horizon.  Hence,
we see that assuming $\ok=0$ reduces error bars on $\omhh$, $\obhh$,
and $\ns(\kfid)$.

{\it Neutrinos and Helium}---Removing $\on h^2$ and $Y_P$ from the
$\ok=0$ case makes further small improvements.  $Y_P$ affects the
ionization history and free electron density so as to change the sound
horizon and damping length.  $\on h^2$ is partially degenerate with
$\omhh$.

{\it Running of the Tilt}---Removing $\alpha$ makes very little
difference except on quantities that depend on extrapolating the
initial power spectrum beyond the well-observed range, e.g.,
$\ns(H_0)$ and $\As$ (see also \cite{Cop98}\ 1998).  As discussed in
\S\,\ref{sec:parameters}, with $\alpha$ free, the spectral tilt becomes
scale-dependent, but there exists a scale at which the error on the
tilt is unchanged by the removal of $\alpha$.  From the fact that the
errors in Table \ref{tab:priors} on $\ns(\kfid)$ change very little as
$\alpha$ is removed, one can infer that $\kfid=0.025\impc$ is close to
this ``pivot point''.  With SDSS information, removing $\alpha$ helps
to better determine $\ts$.

{\it Combined}---For MAP alone, removing tensors on top of removing 
$\ok$, $\on h^2$,
$Y_P$, and $\alpha$ makes a significant difference (compare the
last two columns of the MAP-only section of Table \ref{tab:priors}).
Apparently, combinations of parameters were conspiring to hide their
effects at $\ell<100$.  

In short, stripping the parameter space down to a minimum of baryons,
CDM, cosmological constant, and tilt (with $\tau$ and $\As$ controlling
the normalization) makes a factor of 3 difference in the error bars 
from MAP even for parameters that are not associated with strong degeneracies.  
Including SDSS reduces this dependence; in fact, MAP
plus SDSS in the general parameter space with only curvature fixed  
performs as well as MAP alone in this minimal parameter space!
Degenerate parameters are of course affected much more strongly.

\section{Consistency}\label{sec:consist}

Even if the CMB does end up spinning a seamless tale
of structure formation and cosmological
parameters,
it will not spell the end of cosmology.  Only with stringent
consistency checks from other types of cosmological data can
we be confident that we have eliminated systematic effects in
the data and its analysis.  Moreover, parameter estimation is
only as good as its underlying parameter space; testing CMB
conclusions against other cosmological probes is an important
way to search for unrecognized physical effects.

One of the most important data sets for this task is the galaxy
power spectrum from redshift surveys.  While we have focused in
this paper on the ways in which such surveys can complement
CMB data to improve parameter estimation, this has assumed that 
the measured power spectrum is consistent with the locus of 
allowed models from the CMB.  In fact, with the precision of
upcoming surveys, this is not guaranteed:
there are many possible spectra that will
simply be inconsistent with our understanding of cosmology from
the CMB.  
Attributing the discrepancy to non-linear clustering or
galaxy bias has consequences that are testable within the survey
data; it is not clear that the discrepancy will
be resolved in favor of the CMB.
Explanations involving alterations to the
dark matter sector 
(see e.g.~\cite{Tur97}\ 1997; \cite{Cal98}\ 1998; \cite{Hu98}\ 1998 for 
recent suggestions) 
could severely modify the implications of the CMB for both
low-redshift cosmology and particle physics.

It is also possible that CMB data and the matter power spectrum
will tell a consistent story, while the true nature of the universe
is subtlely otherwise, causing other cosmological measurements to
differ from predictions.  Exotic late-time equations of state
for smooth components are an example of this situation. Therefore,
in the following, we consider consistency checks with other cosmological data sets.

\subsection{$H_0$, $\om$, and Acceleration}

\begin{figure}[tb]
\centerline{\epsfxsize=\colwidth\epsffile{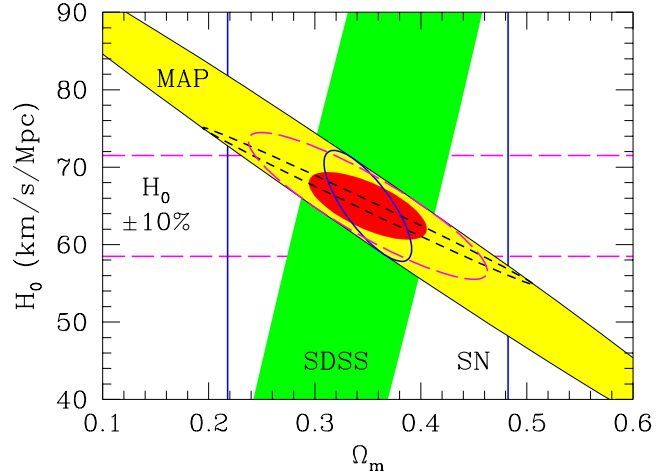}}
\caption{\label{fig:omegahubble}\footnotesize
Constraint regions in the $\om$-$h$ plane from various combinations
of data sets.  
MAP data with polarization yields the ellipse from upper left to lower
right; assuming the universe flat yields a small region (short-dashed line).
SDSS ($\kmax=0.1\ihmpc$) gives the vertical shaded region; combined
with MAP gives the small filled ellipse.
A projection of
future supernovae Ia results (\protect\cite{Teg98a}\ 1998a, middle prediction)
gives the solid vertical lines as bounds; combining this with MAP yields the
solid ellipse.
A direct 10\% measurement of $H_0$ gives the long-dashed lines and ellipse.  
All regions are 68\% confidence.
The fiducial model is the $\om=0.35$ $\Lambda$CDM model.
}\end{figure}

\begin{figure}[tb]
\centerline{\epsfxsize=\colwidth\epsffile{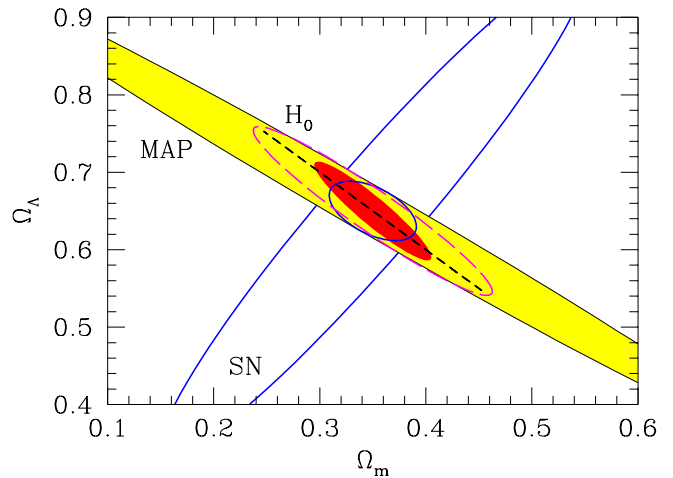}}
\caption{\label{fig:omegalambda}\footnotesize
As Figure \protect\ref{fig:omegahubble}, but for constraints in
the $\om$-$\ol$ plane.
Lines and shadings are unchanged in meaning.
Unlike Figure \protect\ref{fig:omegahubble}, 
assuming the universe flat (short-dashed line) yields a line, not an ellipse.
SDSS-only constraints are not shown.  
}\end{figure}

A rich area for consistency checks is the sector of classical
cosmology: $H_0$, $\om$, $\ol$ (and implicitly $\ok=1-\om-\ol$).  
As explained above,
CMB anisotropies suffer from a severe degeneracy here, but 
a large variety of other precision measurements are available
to clarify the ambiguity and provide consistency checks.  
There are a number of candidates:
the matter power spectrum, supernovae Ia 
(\cite{Per98}\ 1998; \cite{Rie98}\ 1998), and direct measures
of $H_0$ (e.g. \cite{Fre98} 1998; \cite{Bla96} 1996; \cite{Coo98}\ 1998) 
seem particularly promising.
Figures \ref{fig:omegahubble} and \ref{fig:omegalambda}
shows how the combination of each of these 
with CMB data yields a small region in parameter
space.  The overlap of these three regions would be a highly
non-trivial test of cosmology.  Direct measurements of $\om$,
for example from $M/L$ (e.g.\ \cite{Car97b} 1997b) or cluster evolution
(e.g.\ \cite{Car97a} 1997a; \cite{Bah97} 1997),
would also be powerful, and constraints on $\ol$ from gravitational lensing
(e.g.\ \cite{Koc96} 1996) provide a consistency test.

If the smooth component of missing energy is more complicated
than a cosmological constant or curvature, then the constraints
from CMB plus the galaxy power spectrum can still be compared
to results from classical $H_0$ programs.  However, the supernovae
then yield a measurement of this exotic equation of state
(\cite{Gar98}\ 1998; \cite{Hu98b}\ 1998)!

\subsection{Normalization, Reionization, and Bias}\label{sec:biastau}

\begin{figure}[tb]
\centerline{\epsfxsize=\colwidth\epsffile{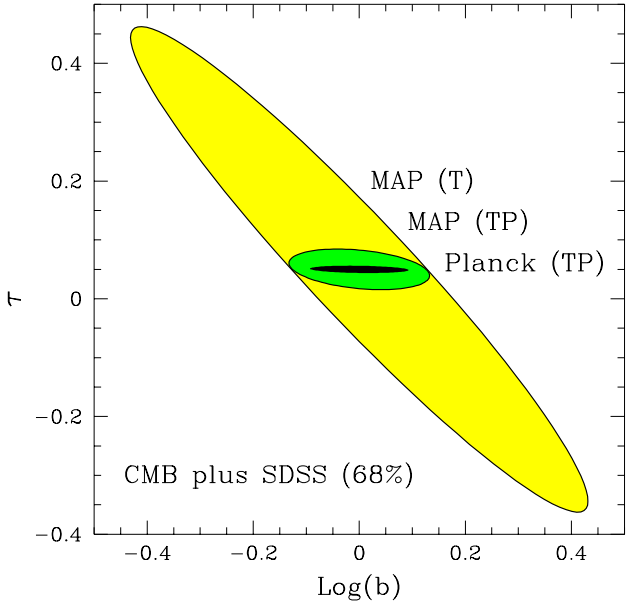}}
\caption{\label{fig:biastau}\footnotesize%
Allowed region in the $b$--$\tau$ plane for the $\Lambda$CDM model.
The 68\% confidence region is shown for MAP+SDSS, with and without
polarization, and Planck+SDSS, with polarization.  $\kmax=0.1\ihmpc$.  
}\end{figure}

Another fruitful area for consistency checks involve parameters 
associated with the amplitude of fluctuations in the CMB and
galaxy power spectrum.          

Several sources of ambiguity in the interpretation of the relative
amplitudes today should be resolved with the larger dynamic range
of upcoming CMB and galaxy data sets.  Tensors affect the
COBE normalization and the spectral tilt adjusts its extrapolation to
smaller scales, but the new satellites will focus on smaller angular
scales for a direct comparison to the scales probed in redshift
surveys. Massive neutrinos affect the matter power spectrum far greater
than the CMB and hence change the relative normalization between these
two. Fortunately, they should also have detectable signatures (see
\S\,\ref{sec:neutrinos}).

Within our parameter space, there are three remaining factors
that alter the observed relative amplitude of the CMB anisotropies and
the galaxy power spectrum.  First, the growth factor between
recombination and the present day depends upon $\ol$ and $\ok$.
Second, galaxy bias alters the normalization of the galaxy power
spectrum relative to that of the matter (and CMB).  
Third, reionization suppresses
CMB anisotropies for a given level of potential fluctuations.

Because the addition of redshift survey information breaks the
angular distance degeneracy, the growth factor will be accurately
known. As discussed in the previous section, there are many 
consistency checks to ensure the validity of this statement.

This leaves the bias and reionization optical depth strongly
covariant, as shown in Figure \ref{fig:biastau}.  Without polarization
information to determine $\tau$ and other dynamical measurements to 
determine $b$, neither will be well-determined.  

With a polarization
detection of the large-angle reionization signal,
however, the error bars on $\tau$ plummet, yielding $b$ and $\sigma_8$
to $\sim\!5\%$ fractionally.  Because polarization measurements are
subject to many systematic effects, consistency checks here are 
particularly important.  With $\Omega_m$ measured precisely, 
any means of constraining $\beta$ or $\sigma_8$ will provide a consistency
check on this measurement, e.g.\ 
redshift distortions (see \cite{Ham97} 1997 for a review), 
peculiar velocity catalogs (\cite{Str95} 1995 and references within), 
and cluster abundances
(\cite{Whi93b}\ 1993b; \cite{Via96}\ 1996; \cite{Eke97}\ 1997; 
\cite{Pen97}\ 1997). 
Additionally, if these measurements can produce limits on $\beta$ 
or below the values given in the Tables ($\simlt$ 5\%), 
then this extra information 
can place leverage on other parameters such as $h$ and $\omhh$ once
$\tau$ has been fixed by the polarization detection.

Conversely, if large-angle CMB foregrounds prevent us from 
extracting the reionization signal, precision measures of $b$ 
or $\sigma_8$ would sharply constrain $\tau$, thereby requiring
our cosmological models to ionize the universe at some particular
redshift.  If this redshift is uncomfortably high or low, it may
challenge our cosmological assumptions, for example, our extrapolation
of $\Pinit(k)$ to yet smaller scales.

\section{Conclusion}\label{sec:concl}

We have presented Fisher matrix calculations of the cosmological
information obtainable with upcoming CMB satellite missions and 
large redshift surveys.  We have used a
considerably larger parameterization of adiabatic CDM than previous
works involving polarization data or redshift surveys.  Within this
space, we have conducted several parameter studies, including
variations in $\ob$, $\on$, and $\ts$.  
In a number of cases, we find softer degeneracies and 
hence larger errors bars than prior work. 
We attribute these discrepancies to artificially broken degeneracies
caused by numerical subtleties such as use of one-sided derivatives and
unfortunate step sizes.

The primary purpose of this paper has been to explore the ways in
which galaxy power spectrum data can provide complementary information
to the CMB, thereby significantly reducing error bars.  In the
language of degeneracies, we seek places in which large-scale structure
offers the means to break CMB degeneracies and vice versa.  The most
important example of this is the angular-diameter distance degeneracy
of the CMB, which causes uncertainty on $h$, $\om$, $\ol$, and 
even $\sigma_8$.  The presence of baryonic oscillations in matter
power spectrum offers a robust way to break this degeneracy.
Redshift-survey data also helps determine the mass of the heaviest
 neutrino,
especially if it exceeds $1\eV$.  Although the projections of the 
Planck satellite with polarization suffice to do most everything
else, less
sensitive CMB experiments can benefit from redshift surveys even
on measuring quantities such as $\omhh$ and $\obhh$.

While CMB and galaxy power spectra themselves offer a myriad of
consistency tests, they also provide a baseline for tests against
other cosmological measurements.  With the detection of the large-angle
polarization signal due to reionization, the two data sets together
yield a $\sim\!5\%$ measure of linear galaxy bias, to be compared to
that obtained from peculiar-velocity and redshift-distortion methods.
Similarly, supernovae distance measurements and Hubble constant 
measurements provide a fertile set of cross-checks on the breaking
of the angular-diameter distance degeneracy.

It is important to remember that the error estimates presented here
reflect the statistical leverage available within these data sets.
We have not attempted to produce a data-analysis pipeline that would
address the horde of obstacles that stand
between the raw data and these cosmological inferences.  
Considerable effort has been channeled toward the development of 
such methods. Foreground removal was mentioned 
in \S\ref{sec:foregrounds}. Pipeline methods have been developed
and implemented for CMB mapmaking 
(\cite{Wri96a}\ 1996; \cite{Wri96b}\ 1996; \cite{Teg97b}\ 1997), 
CMB power spectrum extraction 
(\cite{Teg97Cl}\ 1997c; \cite{Bond98}\ 1998; \cite{Oh98}\ 1998)
and galaxy power spectrum estimation
(\cite{THSVS}\ 1998b). 
All of these are lossless in the sense that they retain all the cosmological 
information quantified by the
Fisher information matrix, but it is likely that
many aspects of the problem will only be
revealed once the data is in hand.  
Approaching the performance
described by analyses such as those in this paper will be one of the 
primary goals (and motivations for) of all this hard work!

In summary, while CMB satellite missions can provide marvelous
cosmological information, they do not make other methods obsolete.  
Additional precision measurements will be needed
both to break the parameter degeneracies of the CMB and to test
for consistency in the face of systematic errors and additional
physical effects.  Only with these precise comparisons can one
build a secure cosmological model.

\bigskip
Acknowledgements: Numerical power spectra were generated with CMBfast 
by Uros Seljak and Matias 
Zaldar\-riaga.\footnote{http://arcturus.mit.edu/$\sim$matiasz/CMBFAST/cmbfast.html}  
We thank Martin White for supplying his results on parameter estimation
for comparison. 
We thank Arthur Kosowsky, Andrew Liddle, David Spergel, Yun Wang, 
Martin White, and Matias Zaldarriaga for useful discussions.
D.J.E.\ is supported by a Frank and Peggy Taplin Membership at the IAS,
W.H.\ by the W.M.\ Keck Foundation,
and D.J.E.\ and W.H.\ by NSF-9513835.  
M.T. was supported by NASA through grant NAG5-6034 and Hubble
Fellowship HF-01084.01-96A from STScI, operated by AURA, Inc.
under NASA contract NAS4-26555.

\appendix
\section{Minimizing errors by reducing covariance}\label{sec:minerror}

At several points in the paper, we are concerned with how the
uncertainties on a particular quantity depend upon the uncertainties
on other related quantities with which it is correlated.  
Here we show that a particular linear combination of the
correlated quantities has the minimum variance and is uncorrelated
with the other parameters.  This variance
is furthermore equal to that of the original quantity under the
assumption that the correlated parameters have been fixed by priors.
We examine optimal combinations of the scalar tilt $\ns$ and running
of the tilt $\alpha$ 
as examples.   The case of the tensor-scalar ratio $T/S$ and tensor
tilt $\nt$  was discussed in \S \ref{sec:tensors}.

\subsection{Proof}

We assume a $n$-dimensional parameter space with independent variables
$p_1$, $\ldots$, $p_n$.
Let the primary quantity be $p_1$ and the $m-1$ quantities ($m\le n$)
that we want to combine be $p_2, \ldots, p_m$.  
We seek a new quantity $X = f(p_1,p_2,\ldots,p_m)$ that has
$\partial f/\partial p_1=1$ and minimum errors.  The former
requirement is to prevent simple rescalings of the variable $p_1$;
$X$ and $p_1$ will in this sense have the same scale.

In the Fisher matrix formalism, all that matters is the gradient of 
$X$ with respect to the independent variables.  
We write $w_j = \partial f/\partial p_j$ and arrange them as a vector $\bfw$; 
by construction, $w_1=1$ and $w_j=0$ for $m<j\le n$.
Then the smallest attainable variance of $X$ will be
\beq
\sigma^2_X = \bfw^T C \bfw,
\eeq
where $C$ is the inverse of the Fisher matrix $F$.
We seek the set of $w_2$, $\ldots$, $w_m$ that minimize $\sigma^2_X$.
Clearly only the submatrix $\tilde{C}$ involving the first $m$ rows and
columns of $C$ can be involved.
Using Lagrange multipliers, one readily shows that the minimum is achieved at 
\beq
w_j=\left\{\begin{array}{ll}
{(\tilde{C}^{-1})_{1j}\over(\tilde{C}^{-1})_{11}} & j\le m, \\
0 & m<j\le n.
\end{array}\right.
\eeq
The minimum value of $\sigma^2_X$ is then $(\tilde{C}^{-1})_{11}^{-1}$.

If one replaces the independent variable $p_1$ by $X$, then the
transformed matrix $C$ will have $C_{1j}=0$ for $j=2,\ldots,m$.
In other words, $X$ is uncorrelated with the variables $p_2,\ldots,p_m$.
This means that $X$ will have the same errors regardless of whether
$p_2,\ldots,p_m$ vary or are held fixed.  If they are held fixed,
then $X$ and $p_1$ are identical.  Hence, the error on $X$
with $p_2,\ldots,p_m$ varying is the same as the error of $p_1$
with them held fixed.

Note that in the case of $m=n$, $\tilde{C}^{-1}=F$.  Hence, $\bfw$ is
just the renormalized first column of the Fisher matrix, and the minimum
variance of $X$ is simply $1/F_{11}$.  Aficionados will recognize this
as the variance of $p_1$ if all other variables are held fixed.

In the case of $m=2$, where we wish to combine $p_1$ with one other variable
$p_2$, we obtain the special case that $X = p_1 - p_2 C_{12}/C_{22}$ and 
$\sigma^2_X = C_{11}-C_{12}^2/C_{22}$.  $X$ and $p_2$ are uncorrelated.

\subsection{A tilt example}

An interesting example of this concerns $\ns$ and $\alpha$.  We
define our independent variable to be $\ns(\kfid)$ and $\alpha$.
For most values of $\kfid$, these two will be correlated.  What
is special in this case is that the minimum-error combination $X$ is
a physically-motivated quantity, namely the tilt $\ns$ at some new scale 
$\kpivot = \kfid \exp(-C_{\ns\alpha}/C_{\alpha\alpha})$
(see eq.\ [\ref{eq:nk}]).
As shown above, $\ns(\kpivot)$ is uncorrelated with $\alpha$ and has
the same error with $\alpha$ varying as $\ns$ on any scale would have
if $\alpha$ were held fixed.

The pivot scale $\kpivot$ at which the error on $\ns(\kpivot)$ is minimized
depends on the experiment and on the fiducial model.  For our $\Lambda$CDM
model and SDSS alone, $\kpivot$ is $0.024\impc$ ($0.088\impc$) for 
$\kmax=0.1\ihmpc$ ($0.2\ihmpc$).  For CMB data (with SDSS to $0.1\ihmpc$
in parentheses), the scales in Mpc$^{-1}$ are 0.034 (0.036) for MAP(T),
0.018 (0.020) for MAP(TP), 0.084 (0.070) for Planck(T), and
0.029 (0.027) for Planck(TP).  As expected, Planck adds more small-scale
sensitivity, while polarization adds more large-scale sensitivity
by resolving the low-$\ell$ degeneracies.  Our choice of $\kfid=0.025\impc$
is close to the desired spot, and the errors grow only to second order 
in $\ln(k)$ away from the minimum:  
\beq
\sigma^2_{\ns(k)} = \sigma^2_{\ns(\kpivot)}+\ln^2(k/\kpivot)\sigma^2_{\alpha}.
\eeq

\section{Numerical Methods}\label{sec:numerics}

\subsection{Derivative Methodology}\label{sec:derivatives}

\begin{table}[tb]\footnotesize
\caption{\label{tab:parameters}}
\begin{center}
{\sc List of Independent Parameters.\\}
\begin{tabular}{rcl}
\tableskip\tableline\tableline\tableskip
Quantity & Step size &Notes \\
\tableskip\tableline\tableskip
$\om h^2$ & $\pm5\%$ \\
$\ob h^2$ & $\pm5\%$ \\
$\on h^2$ & $\pm10\%$ & Equivalent to $m_\nu$\\
$\ol$ & $\pm 0.05\om$ \\
$\ok$ & $\Delta \Omega_m = -0.1\om$ & See \S\ \ref{sec:dafixed} \\
$\tau$ & $\pm 0.02$ \\
$Y_p$ & $\pm 0.02$ & Prior of $\pm0.02$ \\
$\ns(\kfid)$ & $\pm 0.005$ & See eq. (\ref{eq:Pinit})\\
$\alpha$ & $\pm 0.005$ & See eq. (\ref{eq:Pinit})  \\
$\ts$ & Exact \\
$\nt$ & $\pm 0.01$ & Only if $\ts\ne0$. \\
$\As$ & Exact \\
$\beta$ & Exact & $\equiv \om^{0.6}/b$ \\
\tableskip\tableline
\end{tabular}
\end{center}
NOTES.---%
Step sizes are for our $\Lambda$CDM model; those listed as percentages
are fractions of the fiducial value. 
\end{table}

As described in \S \ref{sec:fisher}, 
the calculation of the Fisher matrix reduces to 
manipulation of derivatives of the various power spectra with 
respect to cosmological parameters.  However, the near cancellation
of certain linear combinations of derivatives leaves the Fisher
matrix nearly singular.  As the larger error bars are
themselves inverses of the smaller eigenvalues, it is critical
to prevent numerical effects from perturbing these small eigenvalues.
Hence, constructing and manipulating the derivatives requires care,
lest a parameter degeneracy be broken by numerical effects and yield an
overestimate of the experiment's ability to measure the associated parameters.  
We find that our treatment of certain derivatives reveals
significantly softer directions in parameter space than found by 
previous works (see Appendix \ref{sec:comparison}).
Therefore, we will describe our methods in some detail.

\subsubsection{Two-sided derivatives}

Wherever possible, we take two-sided derivatives.
Writing $f(p)$ for the dependence of 
either $\ln C_\ell$ or $\ln P(k)$ on some parameter
$p$ with all others fixed, this means that
we approximate $f'(p)$ by 
\beq \label{eq:twoside}
\fpt(p)\equiv {f(p+\Dp)-f(p-\Dp)\over 2\Dp}
\eeq
for some small step size $\Dp$.
This is exact to $2^{nd}$ order in $\Dp$, whereas approximating
$f'(p)$ with the one-sided difference
$[f(p+\Dp)-f(p)]/\Dp\approx f'(p+\Dp/2)$
is only good to first order; moreover, the latter
corresponds to an accurate estimate of the derivative
at a slightly shifted parameter value $p+\Dp/2$.  This is critical
when perturbing parameters that change the locations of the Doppler
peaks, because the various derivatives will no longer be in phase! 
If the step size is sufficiently large, these phase shifts will break
the parameter degeneracies.

One also should avoid differencing models that are calculated
with different numerical techniques,
as these can cause discontinuous results as one
adjusts the independent variable.  With CMBfast v2.3.2, 
this situation occurs for geometrically flat versus open models and
for models with differing numbers of species of massive neutrinos.

Because of the desire to use two-sided derivatives, we take non-zero
fiducial values for non-negative quantities such as $\tau=0.05$
and $\on/\om = 0.05$ unless otherwise noted.  Perturbing around $\on\ne0$
also allows us to use one species of massive neutrinos in all cases.

\subsubsection{Derivative step sizes}

What is the best choice of the step size $\Dp$ for
constructing the derivative in 
equation (\ref{eq:twoside})?
We use CMBfast v2.3.2 (including the bug fixes of v2.4.1; 
\cite{Sel96c}\ 1996; \cite{Zal97b}\ 1997) for our numerical derivatives
and find that this version of CMBfast has random numerical noise
at a level of $\sigma\equiv\delta C_\ell/C_\ell \sim 10^{-4}$ 
for geometrically flat models and $10^{-3}$
for open models.\footnote{This does not apply to variations in the initial
power spectrum, as the anisotropies generated by different wavenumbers
are weighted and combined at double precision.  The quoted levels are
for $\ell\lesssim100$ and may increase beyond that.}  
As described by \cite{Pre92}\ (1992), the optimal compromise 
between numerical noise and higher-order Taylor series terms occurs
for $\Dp\sim\sigma^{1/3} p_c$, where $p_c$ indicates the characteristic
scale on which $p$ varies.  Hence, if $p_c\sim p$, 
we should use fractional steps of roughly 5\%, yielding an
accuracy of $\sim\!0.2\%$.  Note that a one-sided derivative with
a 5\% step would yield a truncation-dominated accuracy of $\sim\!5\%$,
which is considerably worse.

Since this estimate is quite crude, we performed a series
of numerical experiments with different step sizes before arriving at
the choices in listed in Table \ref{tab:parameters}. We have
tested that our answers change by less than 10\%
when using half the listed steps.  Convergence is better 
as degeneracies are lifted by complementary information or by
reducing the parameter space via priors.
Only for the open fiducial model do we see some lack of convergence 
due mainly to the larger noise associated with open models.
The step sizes on $\ns$, $\nt$, and $\alpha$ could easily have been
significantly smaller, although reducing them to $10^{-4}$ 
changes the marginalized errors by less than 1\%. 

\subsection{Curvature}\label{sec:curvature}

\begin{figure}[tb]
\centerline{\epsfxsize=\colwidth\epsffile{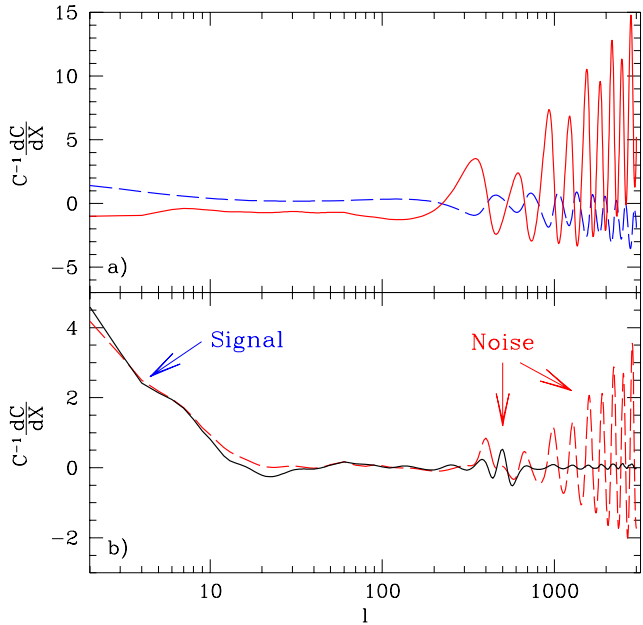}}
\caption{\label{fig:curvderiv}\footnotesize%
(a) Derivatives of $C_{T\ell}$ with respect to $\ok$ at constant $\ol$
(solid line)
and with respect to $\ol$ at constant $\ok$ (dashed line).  
The former is a one-sided
derivative formed by differencing models with $\ok=0.01$ and $\ok=0.003$.
(b) Derivative of $C_{T\ell}$ with respect to $\ok$ at constant $\da$,
constructed in two different ways.  (solid line) result
from differencing two models with $\ol$ and $\ok$ altered so as to keep
$\da$ fixed.  (dashed line) the appropriate linear 
combination of the curves in panel (a); the poor cancellation is due
to the curvature derivative being one-sided and therefore out-of-phase.
The glitch at $\ell\approx500$ in the solid curve is numerical noise
and the reason we need to truncate this derivative.
}\end{figure}

Curvature is known to be strongly degenerate with other effects, such
as a cosmological constant, that alter the late-time evolution of the
universe
(\cite{Hu95}\ 1995 \S VI.B2; \cite{Bon97}\ 1997; \cite{Zal97a}\ 1997). 
Combinations of changes in these parameters that hold the
angular diameter distance to last scattering fixed will leave the
acoustic peaks unchanged (given a proper choice of normalization).
Differences in the decay of potentials at late-times will alter the
large-angle anisotropy; this alteration of the ISW effect is the
primary way to break this degeneracy in the absence of lensing effects 
(\cite{Hu97}\ 1997; \cite{Sto98}\ 1998).

When faced with a strong degeneracy, it is numerically 
desirable to change variables in parameter space 
by introducing a parameter in the degenerate direction.
This has the advantage of containing all numerical
problems in a single (very small) derivative, where
the physical effects can be easily understood and distinguished
from numerical errors making the derivative artificially large.
We adopt this approach for the angular distance degeneracy, as detailed in
the following three subsections.

\subsubsection{Differentiating with $\da$ fixed}
\label{sec:dafixed}
Including variations in curvature around a flat model presents
a particular concern because no closed version of CMBfast is 
available.  Hence, the derivative with respect to curvature is
necessarily a one-sided derivative.   Moreover, the open version
of the code is noisier than the flat, which forces one to take larger
step sizes for the derivatives.  Since adding curvature causes
a significant shift in the location of the CMB peaks, large step
sizes cause the curvature derivative to be out-of-phase with the
$\ol$ derivative.  This is displayed in Figure \ref{fig:curvderiv};
the dashed line in the bottom panel shows that the two derivatives
in the top panel cannot be exactly cancelled.

We work around this problem by altering our coordinate basis so that
the only derivative in a direction with non-zero curvature has 
a convenient physical property.  The derivative of $\ol$ at 
constant $\ok$ may be done in the usual manner.  But for the
derivative of $\ok$ at constant $\ol$, we transform to the basis
of $\ok$ and angular diameter distance $\da$.  The latter is
defined as
\beqa\label{eq:dang}
\da &=& {1\over\sqrt{\ok H_0^2}}
\sinh\left[\sqrt{\ok} \int_0^{z_*} {dz\over \sqrt{E(z)}}\right]\\
E(z) &=& \ol+\ok(1+z)^2+\om(1+z)^3+\Omega_R(1+z)^4 \nonumber,
\eeqa
where $\Omega_R$ is the density of radiation and $z_*$ is the redshift
of recombination (\cite{Hu97a} 1997a).  
We treat massive neutrinos according to the approximation of 
Appendix A of \cite{Hu98a} (1998).
The desired curvature derivative is then
\beq \label{eq:curv}
\left(\partial\over\partial\ok\right)_{\ol} =
\left(\partial\over\partial\ok\right)_{\da} + 
{\left(\partial\da\over\partial\ok\right)_{\ol}  \over
\left(\partial\da\over\partial\ol\right)_{\ok}}
\left(\partial\over\partial\ol\right)_{\ok}.
\eeq
In flat models, the first term is constructed by a one-sided finite 
difference in curvature
corresponding to $\Delta \Omega_m = -0.1\Omega_m$.
The second term consists of a numerical prefactor times the derivative
that we have already calculated.  The first term is special
because with the proper choice of normalization it is a derivative 
in the direction of constant Doppler peak location, shape, and height.
This means that aside from an ISW signal at low $\ell$ and lensing
effects at high $\ell$, the derivative vanishes, allowing one to
isolate the numerical noise.  Also, since the peak structure of both
derivatives enters through the same $\partial/\partial\ol$ at constant
$\ok$ term, the two derivatives are guaranteed to cancel up to 
the treatment of the first term of equation (\ref{eq:curv}).

\subsubsection{Differentiating with fixed high-$z$ normalization}
\label{sec:diffnorm}

To properly implement the technique of the last section, one must
choose a normalization convention that is independent of
$\ol$ and $\ok$.
This condition is satisfied by keeping unchanged the physical situation at
high redshift. It is implemented by holding fixed the 
amplitude of the scalar
gravitational potential at high redshift on some comoving reference scale
$k_{\rm norm}$ in Mpc$^{-1}$
that is outside the horizon at the redshift in question.  
We use the low-redshift
growth function to shift the scalar potential back to the 
pure-matter-domination phase (ignoring radiation).  This produces
the normalization 
\beq\label{eq:normalization}
P(\knorm,z=0) = \As
\left(\knorm\over\kfid\right)
\left(1+ {3\Omega_K H_0^2\over \knorm^2}\right)^2
{H_0^{-4}c^4} \dgr^2;
\eeq
($\ns=1$) in other words, we choose $\As$ as an independent variable and therefore
renormalize the calculated (COBE normalized)  
$P(k)$ and $C_\ell$ by Equation (\ref{eq:normalization})
to hold $\As$ constant when taking all other derivatives.
Here $\dgr$ is the growth function integral (\cite{Pee80} 1980)
\beq\label{eq:dgr}
\dgr = {5\over2}\int^1_0{a^{3/2}\,da\over(\om+\ok a+\ol a^3)^{3/2}}.
\eeq
Note that in the Harrison-Zel'dovich case, the normalization is independent of
$\knorm$ in the large-scale limit; 
we choose a normalization scale of $\knorm^{-1}=3000\mpc$.
If $\ns\ne1$, it is important to pick $\knorm$ to be independent of $h$.
Since the high-redshift ratios of the various types of matter and
radiation are held fixed when perturbing $\ol$ and $\ok$, it doesn't
matter whether we normalize to the matter-domination or
radiation-domination potential.

\subsubsection{Noise clipping}

Even with the derivative technique of the last two sections, 
the numerical noise of CMBfast v2.3.2 breaks the angular diameter
distance degeneracy.  Fortunately, with this choice of derivatives,
the true signal is well-localized in $\ell$.
Neglecting gravitational lensing (see \S \ref{sec:lensing}),
the only signal in the curvature derivative
at constant $\da$ is the low-$\ell$ ISW effect.  This effect scales
as $C_\ell\sim\ell^{\ns-4}$ asymptotically (\cite{Hu96b}\ 1996).  

Since the goal of our analysis is to make robust and conservative
estimates for how accurately cosmological parameters can 
be measured, we assume that no useful cosmological 
information can be extracted from lensing effects or from 
ISW contributions at large $\ell$. 
For $\ell>\ell_{\rm clip}=30$,
we therefore set $C'_\ell = (\ell_{\rm clip}/\ell)^{4-\ns}C'_{\ell_{\rm clip}}$
for this particular derivative, thereby
throwing away all of the high-$\ell$ numerical noise that would otherwise
artificially break the degeneracy above some 
$\ell$-cutoff $\ell_{\rm clip}$.  This numerical noise is
displayed as the solid line in Figure \ref{fig:curvderiv}; the
glitches at $\ell\approx500$ dominate the breaking of
the degeneracy for either satellite.

Hence, while our error bars on $\ol$ and $\ok$ for Planck alone 
(especially with polarization) are overestimated due to the neglect of
lensing,
we protect the softness of this degeneracy against the numerical
problems that would otherwise dominate.

\subsubsection{Effects of $\ok$ and $\ol$ on $P(k)$}

$\ok$ and $\ol$ have degenerate effects on $P(k)$ as well, and so
we must take care in constructing these derivatives.
Fortunately, there is an analytic solution.  
The matter power 
spectrum can be decomposed into an initial power spectrum whose
time dependence reflects only the physics on the largest scales times
a transfer function that incorporates the effects of causal physics
(e.g.\ \cite{Eis98a}\ 1998ab).
$\ok$ and $\ol$ shift the
the $z=0$ normalization of the initial power spectrum relative 
to the level of CMB anisotropies by altering the growth function $\dgr$
[eq.~(\ref{eq:dgr})].
Meanwhile, the transfer function is independent of $\ok$ and $\ol$
if it is measured in real space instead of redshift space.  Hence,
$\ok$ and $\ol$ enter this piece only through their effect on $h$.
We can rewrite this as a derivative with respect to $k$:
\beq\label{eq:dPdcurv}
{d\ln P\over d\ok} =
2{d\ln \dgr\over d\ok} - {1.2\over\om}
+ {1\over2\om}\left({d\ln P\over d\ln k}-1\right).
\eeq
This avoids having to difference two different Boltzmann code outputs
and allows us to track the oscillations in the power spectrum and
its derivative more accurately.
The middle term in equation (\ref{eq:dPdcurv}) comes from the
derivative of $\ln b$, since $\beta$ is our independent variable.
The derivative with respect to $\ol$ has an equivalent formula.
The assumption that the transfer function in real space is 
independent of $\ok$ and $\ol$ is actually violated slightly at 
small scales if massive neutrinos are important due to differences
in the infall of the neutrinos (\cite{Hu98a} 1998);
we ignore this effect as it is tiny and primarily beyond the linear regime.

\subsection{$\tau$ and Normalization}

Reionization mimics a suppression of the amplitude of the primary CMB anisotropies
on all but the largest scales.  For small optical depths, the secondary fluctuations
generated by these late-time scatterings are small because only a small
fraction of photons are affected and because the scattering occurs over
a sufficient range of distances along the line of sight that
small-scale perturbations are averaged out (\cite{Kai84} 1984).  Hence, for $\tau\ll 1$,
the main effect is a suppression of $C_\ell$ by $\exp(-2\tau)$ for all
but the lowest $\ell$.  If the normalization of the primary
fluctuations is unknown, then we can hide the reionization simply by
increasing their normalization by a corresponding amount.  The
resulting degeneracy is difficult to break in the temperature
anisotropy data because the rise at low $\ell$ up to the original
fluctuation level is hidden by cosmic variance.  With polarization
data, reionization can be more easily separated because it produces a
bump at large-angles whereas fluctuations at high redshift
cannot produce much large-angle polarization due to causality
(\cite{Kai83} 1983).

With this level of degeneracy, one should worry about artificially
breaking the cancellation between the $\tau$ and $\As$ derivatives.
However, we find that CMBfast is sufficiently accurate to track this
degeneracy without invoking tricks similar to those used for the
curvature, e.g., differentiating with 
$A_S e^{-\tau}$ held fixed.

Note that if our chosen normalization were the COBE normalization,
we would also have to worry about degeneracies between the 
tensor-to-scalar ratio $\ts$ and the scalar normalization.  
By normalizing to the scalar potential fluctuations, however,
we avoid this problem.

\section{Comparison to Previous Work}\label{sec:comparison}

The results in the main part of this paper are not directly comparable
to those of previous papers due to differences in parameter spaces or
fiducial models.  In this Appendix, we attempt to provide as direct a
comparison as we can for some of the results presented in these papers.
We have in all cases matched the quoted parameter spaces and experimental
specifications.
Except in the case of \cite{Whi98} (1998), we have not taken care to 
provide an exact match to the normalization
in these comparisons.  All groups are normalizing to the
COBE value for the signal to noise in equation (\ref{eq:cmbcov}).
While normalization differences of a few percent are possible, these
would not affect the parameter results beyond this small level.

\begin{table}[tb]\footnotesize
\caption{\label{tab:compbet}}
\begin{center}
{\sc Comparison to Bond \etal\ (1997)\\}
\begin{tabular}{lcc\colskip cc}
\tableskip\tableline\tableline\tableskip
& \multicolumn{2}{c\colskip}{MAP} & \multicolumn{2}{c\colskip}{Planck (HFI)} \\
Quantity & BET & EHT & BET & EHT \\
\tableskip\tableline\tableskip
$2h$ & 0.19 & 0.40 & 0.02 & 0.08\\
$4\ol h^2$ & 0.49 & 1.2 & 0.05 & 0.20 \\
$\ln\obhh$ & 0.09 & 0.14 & 0.006 & 0.018 \\
$4\omhh$ & 0.18 & 0.38 & 0.02 & 0.04 \\
$4\on h^2$ & 0.07 & 0.09 & 0.02 & 0.03 \\
$\ns$ & 0.06 & 0.10 & 0.006 & 0.015 \\
$\ts$ & 0.38 & 0.69 & 0.09 & 0.11 \\
$\tau$ & 0.22 & 0.34 & 0.16 & 0.30 \\
$\ln\sigma_8$ & 0.28 & 0.57 & 0.18 & 0.33 \\
\tableskip\tableline
\end{tabular}
\end{center}
NOTES.---%
Fiducial model has $\om=1$, $h=0.5$, $\ob=0.05$, $\on=0$,
$\ol=0$, $\tau=0$, $\ns=1$, $\ts=0$, and $Y_P=0.23$ (with
a prior of $\Delta Y_P = 0.02$.
All errors are $1-\sigma$.  
BET columns contain the errors quoted in \protect\cite{Bon97}\ (1997);
EHT columns contain the errors we find using their experimental 
specifications for MAP and the High-Frequency Instrument (HFI) of Planck.
All columns are temperature data only. 
\end{table}

In Table \ref{tab:compbet}, we compare marginalized errors with
\cite{Bon97}\ (1997) for their standard CDM model.  We use their
experimental specifications for MAP and the 
Planck High-Frequency Instrument HFI
(temperature only) and restrict ourselves to their parameter space
($\ok=\alpha=0$).  Note that we have by necessity used one-sided
derivatives for $\tau$ and $\on$ differencing models with
$0.01$ and $0.001$ in each parameter.  Because
the curvature is not varied, the tricks of \S \ref{sec:curvature}
are not needed.  We find significantly 
(up to a factor of 3) larger errors on some parameters.

\begin{table}[tb]\footnotesize
\caption{\label{tab:compzss}}
\begin{center}
{\sc Comparison to Zaldarriaga \etal\ (1997)\\}
\begin{tabular}{lcc\colskip cc}
\tableskip\tableline\tableline\tableskip
& \multicolumn{2}{c\colskip}{MAP(T)} & \multicolumn{2}{c\colskip}{MAP(TP)} \\
Quantity & ZSS & EHT & ZSS & EHT \\
\tableskip\tableline\tableskip
$h$     & 0.092  & 0.17   & 0.051   & 0.040 \\
$\ol$   & 0.53   & 1.1    & 0.29    & 0.25 \\
$\obhh$ & 0.0010 & 0.0019 & 0.00061 & 0.00055 \\
$\tau$  & 0.13   & 0.21   & 0.021   & 0.022 \\
$\ns$   & 0.059  & 0.12   & 0.031   & 0.028 \\
$\ts$   & 0.39   & 0.77   & 0.22    & 0.19 \\
\tableskip\tableline\tableskip
$h$     & 0.017   & 0.025   & 0.016    & 0.017 \\
$\ol$   & 0.098   & 0.15   & 0.093    & 0.103 \\
$\obhh$ & 0.00030 & 0.00043 & 0.00028  & 0.00035 \\
$\tau$  & 0.12    & 0.15   & 0.021    & 0.021 \\
$\ns$   & 0.0098  & 0.021   & 0.0048   & 0.010 \\
$\ts$   & \nodata & \nodata & \nodata & \nodata \\
\tableskip\tableline
\end{tabular}
\end{center}
NOTES.---%
Fiducial model has $\om=1$, $h=0.5$, $\ob=0.05$, $\tau=0.05$, $\ns=1$,
and $\ts=0$.  $\ok$, $\on$, $Y_P$, and $\alpha$ are held fixed.
The top set of numbers is with $\ts$ allowed to vary; the bottom set is with
$\ts$ fixed.
All errors are $1-\sigma$.  
ZSS columns contain the errors quoted in \protect\cite{Zal97a}\ (1997);
EHT columns contain the errors we find using their experimental 
specifications for MAP with and without polarization.
\end{table}

\cite{Zal97a}\ (1997) analyzed a standard CDM model with a smaller
parameter space.  Restricting to this space ($\ok=\on=\alpha=\Delta
Y_P=0$) and adopting their specifications for MAP yields the results in
Table \ref{tab:compzss}.  Because $\tau\ne0$ and massive neutrinos are
excluded, no one-sided derivatives are required.  The results agree
well with temperature and polarization data (except for $\ns$)
but can be up to a factor of 2 different with temperature data alone.

\begin{table}[tb]\footnotesize
\caption{\label{tab:compwss}}
\begin{center}
{\sc Comparison to Wang \etal\ (1998)\\}
\begin{tabular}{rlcc\colskip cc}
\tableskip\tableline\tableline\tableskip
& & \multicolumn{2}{c\colskip}{MAP(T)} & \multicolumn{2}{c\colskip}{MAP(TP)} \\
Model & Quantity & WSS & EHT & WSS & EHT \\
\tableskip\tableline\tableskip
SCDM 
& $\ln h$    & 0.052  & 0.051  & 0.033  & 0.033 \\
& $\ol$      & 0.15  & 0.15   & 0.091  & 0.097 \\
& $\ln\obhh$ & 0.028  & 0.031  & 0.020  & 0.024 \\
& $\ln\tau$   & 2.4    & 2.9    & 0.39   & 0.38 \\
& $\ns$      & 0.017 & 0.020  & 0.0085 & 0.0090 \\
\tableskip\tableline\tableskip
$\Lambda$CDM 
& $\ln h$    & 0.066  & 0.077  & 0.032  & 0.035 \\
& $\ln \ol$  & 0.076  & 0.089  & 0.037  & 0.043 \\
& $\ln\obhh$ & 0.044  & 0.052  & 0.021  & 0.023 \\
& $\ln\tau$   & 1.3   & 1.6    & 0.18   & 0.18 \\
& $\ns$      & 0.035 & 0.041  & 0.014 & 0.014 \\
\tableskip\tableline
\end{tabular}
\end{center}
NOTES.---%
Top set of numbers are for a standard CDM fiducial model with 
$\om=1$, $h=0.5$, $\ob=0.05$, $\tau=0.05$, and $\ns=1$.
Bottom set are for a $\Lambda$CDM model with
$\om=0.3$, $h=0.65$, $\ob=0.06$, $\ol=0.7$, $\tau=0.1$, and $\ns=1$.  
$\ok$, $\on$, $Y_P$, $\ts$, and $\alpha$ are held fixed.
All errors are $1-\sigma$.  
WSS columns contain the errors quoted in the revised 
version of \protect\cite{Wan98}\ (1998; private communication).
EHT columns contain the errors we find using their experimental 
specifications for MAP with and without polarization.
\end{table}

We suspect that the numerical issues addressed in Appendix
\ref{sec:derivatives} are responsible for discrepancies described above.
Unfortunately, based on the published information, we cannot confirm in
the above cases that this is the source of the discrepancies.

Our comparison with \cite{Wan98}\ (1998) demonstrates the situation.
We compared two fiducial models, adopting
their parameter space ($\ok=\alpha=\on=\ts=\Delta Y_P=0$) and
specifications for MAP (with $\fsky=0.8$).  At first, agreement
was poor, with their errors being as much as a factor of 5 smaller
on one model.  One-sided derivatives had been used; when
\cite{Wan98}\ kindly recomputed their results using two-sided
derivatives and smaller steps on certain parameters, the
agreement became quite good ($\lesssim20\%$), 
as show in Table \ref{tab:compwss}.  
We did not compare results for SDSS and CMB together because the
treatments are quite different: they remove baryon oscillations from
$P(k)$ and apply a non-linear evolution correction.

In another situation in which we could confirm that
two-sided derivatives were used, we have attained even better agreement.
We have compared a $\Lambda$CDM model with the preliminary results
of \cite{Whi98} (1998), who uses a hierarchy Boltzmann code rather
than CMBfast.
Under a parameter space of $\obhh$, $\omhh$, $\ol$,
$\tau$, $\ns$, and $\As$, the marginalized error bars agree to 3\% 
with and without
polarization for MAP and Planck.  Adding $\ok$ causes larger
discrepancies, mostly along the direction of the angular diameter
distance degeneracy; we suspect the hierarchy code is more stable for
$\ok\ne0$, but the differences should disappear once the degeneracy is
broken by outside information.

Degeneracies are quite important even in these smaller parameter 
spaces.  As an example, consider the SCDM model from 
\cite{Zal97a}\ (1997).  Holding $\tau$ fixed removes the 
reionization-normalization degeneracy and leaves $\omhh$, $\obhh$,
$\ol$, $\ns$, $\ts$, and $\As$ to vary.  This set of parameters
in this fiducial model can be combined to cancel the normalization
temperature derivative to better than 1 part in 100 over the 
$\ell$-range $100-800$, mostly through a combination of tilt, tensors,
and $\omhh$.  
Hence, even in this small space, one needs to control the
derivatives to high numerical accuracy.  
For example, an 0.05 one-sided
step in tilt pivoting around the Hubble distance produces a 16\%
derivative error at $\ell\approx600$ due to second-order terms.
These considerations and the comparison with \cite{Wan98}\ (1998)
and \cite{Whi98} (1998) lead us to conclude that two-sided derivatives
are crucial for achieving accurate answers with present-day codes.


\begin{thebibliography}{99}\frenchspacing

\bibitem[Bahcall \etal]{Bah97}
    Bahcall, N.A., Fan, X., \& Cen, R.\ 1997, \apj, 485, L53

\bibitem[Bersanelli \etal]{Ber96}
    Bersanelli M. {\etal} 1996, {\it COBRAS/SAMBA, Phase A Study for an ESA
M3 Mission}, ESA Report D/SCI(96)3

\bibitem[Blanchard \& Schneider]{Bla87}
    Blanchard, A., \& Schneider, J. 1987, A\&A, 184, 1

\bibitem[Blandford \& Kundic]{Bla96}
    Blandford, R.D., \& Kundic, T. 1996, in The Extragalactic Distance Scale,
    eds. M. Livio, M. Donahue, \& N. Panagia (Cambridge Univ. Press: Cambridge)
    [astro-ph/9611229]

\bibitem[Blumenthal \etal]{Blu84}
    Blumenthal, G. et al. 1984, Nature, 311, 517

\bibitem[Bond \& Szalay]{Bon83}
    Bond, J.R., \& Szalay, A.S. 1983, \apj, 274, 443
  
\bibitem[Bond \etal]{Bon94}
    Bond, J.R., Crittenden, R., Davis, R.L., Efstathiou, G., \&
    Steinhardt, P.J. 1994, \prl, 72, 13

\bibitem[Bond \& Jaffe]{Bon96}
    Bond, J.R., \& Jaffe, A.H. 1997, in Microwave Background Anisotropies,
    ed. Bouchet, F., \etal\ (Editions Frontieres: Singapore)
    [astro-ph/9610091]

\bibitem[Bond \etal]{Bon97}
    Bond, J.R., Efstathiou, G., \& Tegmark, M. 1997, \mnras, 291, L33

\bibitem[Bond \etal]{Bond98}
Bond, J.R., Jaffe, A.H. \& Knox, L. 1998, \prd, 2117, 1998

\bibitem[Brandt \etal]{Bra94}
    Brandt W.N. \etal~1994, \apj, 424, 1 


\bibitem[Bunn \& White]{Bun97}
    Bunn, E.F., \& White, M. 1997, \apj, 480, 6

\bibitem[Caldwell \etal]{Cal98}
    Caldwell, R.R., Dave, R., \& Steinhardt, P.J., 1998, \prl, 80, 1582

\bibitem[Carlberg \etal]{Car97a}
    Carlberg, R.G., Morris, S.L., Yee, H.K.C., \& Ellingson, E.\ 1997a,
    \apj, 479, L19

\bibitem[Carlberg \etal]{Car97b}
    Carlberg, R.G., Yee, H.K.C., \& Ellingson, E.\ 1997b, \apj, 478, 462

\bibitem[Cole \& Efstathiou]{Col89}
    Cole, S., \& Efstathiou, G. 1989, \mnras, 239, 195

\bibitem[Coles]{Col93}
    Coles, P. 1993, \mnras, 262, 1065

\bibitem[Cooray et al.]{Coo98}
    Cooray, A.R., Carlstrom, J.E., Joy, M. Grego, L. Holzapfel, W. \& Patel, S.K. 1998 
    [astro-ph/9804149]   

\bibitem[Copeland \etal]{Cop98}
    Copeland, E.J., Grivell, I.J., \& Liddle, A.R. 1998, \mnras, in press
    [astro-ph/9712028]

\bibitem[David \etal]{Dav95}
    David, L.P., Jones, C., \& Forman, W.\ 1995, \apj, 445, 578

\bibitem[Davis \etal]{Dav92}
    Davis, R.L., Hodges, H.M., Smoot, G.F., Steinhardt, P.J., \& Turner, M.S.
                1992, \prl, 69, 1856
\bibitem[Dodelson \etal]{Dod96}
    Dodelson, S., Gates, E., \& Stebbins, A.\ 1996, \apj, 467, 10

\bibitem[Dodelson, Gates \& Turner]{Dod96b}
    Dodelson, S., Gates, E., \& Turner, M.S.\ 1996, Science, 274, 69

\bibitem[Efstathiou \& Bond]{Efs98}
    Efstathiou, G., \& Bond, J.R. 1998, \mnras, submitted

\bibitem[Eisenstein \& Hu]{Eis98a}
    Eisenstein, D.J., \& Hu, W. 1998a, \apj, 496, 605

\bibitem[Eisenstein \& Hu]{Eis98b}
    Eisenstein, D.J., \& Hu, W. 1998b, \apj, submitted [astro-ph/9710252]

\bibitem[EHT]{EHT}
    Eisenstein, D.J., Hu, W., \& Tegmark, M. 1998, \apjl, in press (EHT)
    [astro-ph/9805239]

\bibitem[Eke \etal]{Eke97}
    Eke, V.R., Cole, S., \& Frenk, C.S. 1996, \mnras, 282, 263

\bibitem[Evrard]{Evr97}
    Evrard, A.E. 1997, \mnras, 292, 289

\bibitem[Freedman et al.]{Fre98}
    Freedman, W.L., Mould, J.R., Kennicut, R.C., \& Madore, B.F. 1998,
    IAU Symposium 183 [astro-ph/9801080]

\bibitem[Fry \& Gazta\~naga]{Fry93}
    Fry, J.N., \& Gazta\~naga, E. 1993, \apj, 413, 447

\bibitem[Garnavich \etal]{Gar98}
    Garnavich, P., \etal\ 1998, \apj, in press
    [astro-ph/9806396]

\bibitem[Gawiser \& Silk]{Gaw98}
    Gawiser, E., \& Silk, J. 1998, Science, 280, 1405

\bibitem[Goldberg \& Strauss]{Gol98}
    Goldberg, D.M., \& Strauss, M. 1998, \apj, 495, 29

\bibitem[Guiderdoni \etal]{Gui98}
    Guiderdoni, B.,  Hivon, E., Bouchet, F. R., \& Maffei, B. 1998, 
    \mnras, 295, 877
		     
\bibitem[Hamilton]{Ham97}
    Hamilton, A.J.S. 1997, in Ringberg Workshop on Large-Scale Structure,
    ed. Hamilton, D. (Kluwer Academic: Dordrecht)
    [astro-ph/9708102]

\bibitem[Hatton \& Cole]{Hat98}
    Hatton, S.J., \& Cole, S. 1998, MNRAS, 296, 10

\bibitem[Heavens \& Taylor]{Hea97}
Heavens, A.F. \& Taylor, A.N. 1997, \mnras, 290, 456
       
\bibitem[Hobson \etal]{Hob98}
    Hobson, M.P., Jones, A.W., Lasenby, A.N., \& Bouchet, F.R. 1998,
    \mnras, in press
    [astro-ph/9806387]

\bibitem[Holtzman]{Hol89}
    Holtzman, J.A., 1989, \apj S, 71, 1 

\bibitem[Hu]{Hu98}
    Hu, W., 1998, \apj (in press) [astro-ph/9801234]

\bibitem[Hu \& Sugiyama]{Hu95}
    Hu, W., \& Sugiyama, N. 1995, \prd, 51, 2599

\bibitem[Hu \& Sugiyama]{Hu96}
    Hu, W., \& Sugiyama, N.  1996, \apj, 471, 542

\bibitem[Hu \etal]{Hu97}
    Hu, W., Sugiyama, N., \& Silk, J. 1997, \nat, 386, 37

\bibitem[Hu \& White]{Hu96b}
    Hu, W. \& White, 1996, M. A\& A, 315, 33

\bibitem[Hu \& White]{Hu97a}
    Hu, W., \& White, M. 1997a, \apj, 479, 568

\bibitem[Hu \& White]{Hu97p}
    Hu, W., \& White, M. 1997b, NewA, 2, 323

\bibitem[Hu \& Eisenstein]{Hu98a}
      Hu, W., \& Eisenstein, D.J., 1998, \apj, 498, 497

\bibitem[HET]{HET}
      Hu, W., Eisenstein, D.J., \& Tegmark, M. 1998a, \prl, 80, 5255 (HET)

\bibitem[Hu \etal]{Hu98b}
      Hu, W., Eisenstein, D.J., Tegmark, M., \& White, M. 1998b, \prd, 
	submitted [astro-ph/9806362]

\bibitem[Huey \etal]{Hue98}
    Huey, G., Wang, L., Dave, R., Caldwell, R.R., Steinhardt, P.J.
    1998, preprint [astro-ph/9804285]

\bibitem[Jain \& Bertschinger]{Jai94}
    Jain, B., \& Bertschinger, E.\ 1994, \apj, 431, 495

\bibitem[Jungman \etal]{Jun96a}
    Jungman, G.,  Kamionkowski, M., Kosowsky, A. \& Spergel, D.N. 1996a,
    \prl, 76, 1007

\bibitem[Jungman \etal]{Jun96b}
    Jungman, G.,  Kamionkowski, M., Kosowsky, A. \& Spergel, D.N. 1996b,
    \prd, 54, 1332

\bibitem[Kaiser]{Kai83}
    Kaiser, N. 1983, \mnras, 202, 1169

\bibitem[Kaiser]{Kai84}
    Kaiser, N. 1984, \apj, 282, 374

\bibitem[Kaiser \& Peacock]{Kai91}
    Kaiser, N., \& Peacock, J.A. 1991, \apj, 379, 482

\bibitem[Kamionkowski \etal]{Kam97}
    Kamionkowski, M., Kosowsky, A., \& Stebbins, A. 1997, \prl, 78, 2058

\bibitem[Kamionkowski \& Kosowsky]{Kos98}
    Kamionkowski, M., \& Kosowsky, A. 1998, \prd, 57, 685

\bibitem[Keating \etal]{Kea98}
    Keating, B., Timbie, P., Polnarev, A., \& Steinberger, J.  1998, \apj, 495, 580

\bibitem[Kinney]{Kin98}
    Kinney, W.H., preprint, astro-ph/9806259

\bibitem[Knox]{Kno95}
    Knox, L. 1995, \prd, 52, 4307

\bibitem[Kochanek]{Koc96}
    Kochanek, C.S. 1996, \apj, 466, 638


\bibitem[Liddle \& Lyth]{Lid93}
    Liddle, A.R. \& Lyth, D.H. 1993, Phys. Rep., 231,  1

\bibitem[Lineweaver]{Lin98}
    Lineweaver, C.H. 1998, preprint, astro-ph/9805326

\bibitem[Lyth]{Lyt97}
    Lyth, D.H. 1997, \prl, 78, 1861

\bibitem[Ma \& Bertschinger]{Ma95}
    Ma, C.P., \& Bertschinger, E. 1995, \apj, 455, 7

\bibitem[Mann \etal]{Man98}
    Mann, R.G., Peacock, J.A., \& Heavens, A.F. 1998, \mnras, 293, 209

\bibitem[Meiksin \etal]{Mei98}
    Meiksin, A., Peacock, J.A., \& White, M. 1998, in preparation

\bibitem[Metcalf \& Silk]{Met97}
    Metcalf, R.B., \& Silk, J. 1997, preprint 
    [astro-ph/9708059]

\bibitem[Oh \etal]{Oh98}
Oh, S.P., Spergel, D.N. \& Hinshaw, G. 1998, preprint
[astro-ph/9805339]

\bibitem[Peacock \& Dodds]{Pea94}
    Peacock, J.A., \& Dodds, S.J. 1994, \mnras, 267, 1020

\bibitem[Peacock]{Pea97}
    Peacock, J.A. 1997, \mnras, 284, 885

\bibitem[Peebles \& Yu]{Pee70}
    Peebles, P.J.E. \& Yu, J.T. 1970, \apj, 162, 815

\bibitem[Peebles]{Pee80}
    Peebles, P.J.E. 1980, The Large-Scale Structure of the Universe
    (Princeton: Princeton Univ. Press)

\bibitem[Pen]{Pen97}
    Pen, U.-L. 1997, preprint
    [astro-ph/9610147]

\bibitem[Perlmutter \etal]{Per98}
    Perlmutter, S., \etal\ 1998, \nat, 391, 51

\bibitem[Press \etal]{Pre92}
    Press, W.H., Teukolsky, S.A., Vetterling, W.T., \& Flannery, B.P.
    Numerical Recipes, 2nd ed. (Cambridge Univ: Cambridge)

\bibitem[Refregier \etal]{Ref98}
    Refregier, A., Spergel, D.N., \& Herbig, T. 1998, preprint
    [astro-ph/9806349]

\bibitem[Riess \etal]{Rie98}
    Riess, A.G., \etal\ 1998, \aj, in press
    [astro-ph/9805201]

\bibitem[Scherrer \& Weinberg]{Sch98}
    Scherrer, R.J., \& Weinberg, D.H. 1997, preprint [astro-ph/9712192]

\bibitem[Schramm \& Turner]{Schr98}
    Schramm, D.N., \& Turner, M.S. 1998, Rev. Mod. Phys., 70, 303

\bibitem[Scott \etal]{Sco93}
    Scott, D., Srednicki, M., \& White, M. 1994, \apj, 421, L5

\bibitem[Scott \etal]{Sco95}
    Scott, D., Silk, J., \& White, M. 1995, Science, 268, 829

\bibitem[Seljak]{Sel96a}
    Seljak, U. 1996a, \apj, 463, 1

\bibitem[Seljak]{Sel96b}
    Seljak, U. 1996b, \apj, 482, 6

\bibitem[Seljak \& Zaldarriaga]{Sel96c}
    Seljak, U., \& Zaldarriaga, M. 1996, \apj, 469, 437

\bibitem[Stompor \& Efstathiou]{Sto98}
    Stompor, R., \& Efstathiou, G. 1998, \mnras, submitted 
    [astro-ph/9805294]

\bibitem[Strauss \& Willick]{Str95}
    Strauss, M., \& Willick, J. 1995, Phys. Rep., 261, 271

\bibitem[Sunyaev \& Zel'dovich]{Sun70}
    Sunyaev, R., \& Zel'dovich, Ya.B. 1970, \apss, 7, 3

\bibitem[Tegmark]{Teg97a}
    Tegmark, M. 1997a, \prl, 79, 3806

\bibitem[Tegmark]{Teg97b}
    Tegmark, M. 1997b, \prd, 56, 4514

\bibitem[Tegmark]{Teg97Cl}
Tegmark, M. 1997c, \prd, 55, 5895

\bibitem[Tegmark \& Efstathiou]{TE96}
    Tegmark, M. \& Efstathiou, G. 1996, \mnras, 281, 1292

\bibitem[Tegmark \etal]{Teg98a}
    Tegmark, M., Eisenstein, D. J., Hu, W., \& Kron, R. 1998a, \apjl, submitted
    [astro-ph/9805117]

 
\bibitem[Tegmark \etal]{THSVS}
    Tegmark, M., Hamilton, A.J.S., Strauss, M.A., Vogeley, M.A. \& Szalay, A.S. 1998b,
    \apj, 499, 555

       
\bibitem[Tegmark]{Teg98b}
    Tegmark, M. 1998, \apj, 502, 1

\bibitem[Tegmark \& de Oliveira-Costa]{TegOli98}
    Tegmark, M. \& de Oliveira-Costa, A. 1998, \apjl, 500, 83	

\bibitem[Tegmark \etal]{Teg97c}
    Tegmark, M., Taylor, A.N., \& Heavens, A.F. 1997, \apj, 480, 22

\bibitem[Toffolatti \etal] {Tof98}
    Toffolatti, L., Argueso Gomez, F., De Zotti, G., Mazzei, P., 
    Franceschini, A., Danese, L., Burigana, C. 1998, \mnras, 297, 11

       
\bibitem[Turner \& White]{Tur96}
    Turner, M.S., \& White, M. 1996, \prd, 53, 6822

\bibitem[Turner \& White]{Tur97}
    Turner, M.S., \& White, M. 1997, \prd, 56, 4439

\bibitem[Viana \& Liddle]{Via96}
    Viana, P.T.P., \& Liddle, A.R. 1996, \mnras, 281, 323

\bibitem[Wang \etal]{Wan98}
    Wang, Y., Spergel, D.N., Strauss, M.A. 1998, \apj, in press
    [astro-ph/9802231]

\bibitem[Webster \etal]{Web98}
    Webster, M., Hobson, M.P., Lasenby, A.N., Hahav, O., \& Rocha, G.
    1998, \apjl, submitted
    [astro-ph/9802109]

\bibitem[Weinberg]{Wei95}
    Weinberg, D.H. 1995, in Wide-Field Spectroscopy and the Distant Universe,
    eds. S.J. Maddox and A. Arag\'on-Salamanca 
    (Singapore: World Scientific), 129

\bibitem[Weinberg et al.]{Wei97}
    Weinberg, D.H., Miralda-Escud\'e, J., Hernquist, L, \& Katz, N.
    1997, \apj, 490, 564

\bibitem[White \& Fabian]{Whi95}
    White, D.A., \& Fabian, A.C.\ 1995, \mnras, 273, 72

\bibitem[White \etal]{Whi93b}
    White, S.D.M., Efstathiou, G., \& Frenk, C.S. 1993, \mnras, 262, 1023

\bibitem[White \etal]{Whi93a}
    White, S.D.M., Navarro, J.F., Evrard, A.E., \& Frenk, C.S. 1993a,
    \nat, 366, 429

\bibitem[White]{Whi96}
    White, M. 1996, \prd, 53, 3011

\bibitem[White]{Whi98}
    White, M. 1998, in preparation

\bibitem[Wright]{Wri96a}
    Wright, E.L., Hinshaw, G., Bennett, C.L. 1996, \apjl, 458, L53

\bibitem[Wright]{Wri96b}
Wright, E.L. 1996, preprint 
[astro-ph/9612006]

\bibitem[Zaldarriaga \& Seljak]{Zal97b}
    Zaldarriaga, M., \& Seljak, U. 1997, \prd, 55, 1830

\bibitem[Zaldarriaga \etal]{Zal97a}
    Zaldarriaga, M., Spergel, D.N., \& Seljak, U. 1997, \apj, 488, 1

\nonfrenchspacing
\end{thebibliography}
\end{document}